\documentclass[graybox,natbib,nosecnum]{svmult}
\bibpunct{(}{)}{;}{a}{}{,} 

\pdfoutput=1   

\usepackage{mathptmx}       
\usepackage{helvet}         
\usepackage{courier}        
\usepackage{type1cm}        

\usepackage{makeidx}         
\usepackage{graphicx}        
\usepackage{multicol}        
\usepackage[bottom]{footmisc}
\usepackage[normalem]{ulem}	
\usepackage{hyperref}  
\usepackage[margin=10pt]{subfig}
\usepackage{soul}   
\usepackage{textcomp}

\newcommand{\ignore}[1]{}

\makeindex

\begin{document}

\title*{Connecting Planetary Composition with Formation}
\author{Ralph E. Pudritz, Alex J. Cridland, \& Matthew Alessi}
\institute{Ralph E. Pudritz \at Department of Physics and Astronomy, McMaster University, Hamilton, Ontario, Canada, L8S 4E8; Origins Institute, McMaster University, Hamilton, Ontario, Canada, L8S 4E8,
\email{pudritz@mcmaster.ca}
\and Alex J. Cridland \at  Leiden Observatory, Leiden University, 2300 RA Leiden, the Netherlands  \email{cridland@strw.leidenuniv.nl  }
\and Matthew Alessi \at Department of Physics and Astronomy, McMaster University, Hamilton, Ontario, Canada, L8S 4E8, \email{alessimj@mcmaster.ca}}

\maketitle

\abstract{  The rapid advances in observations of the  different populations of exoplanets,  the characterization of their host stars and the 
links to the properties of their planetary systems, the detailed studies of protoplanetary disks, and the experimental study of the interiors and
composition of the massive planets in our solar system provide a firm basis for the next big question in planet formation theory.  
How do the elemental and chemical compositions of planets connect with their formation?  The answer to this requires that the various pieces of 
planet formation theory be linked together in an end-to-end picture that is capable of addressing these large data sets.   In this review,
we discuss the critical elements of such a picture and how they affect the chemical and elemental make up of forming planets.  Important
issues here include the initial state of forming and evolving disks, chemical and dust processes within them, the migration of planets and
the importance of planet traps, the nature of angular momentum transport processes involving turbulence and/or MHD disk winds, planet formation theory,
and advanced treatments of  disk astrochemistry.  All of these issues affect, and are affected by the chemistry of disks which is driven by X-ray ionization of the 
host stars.  We discuss how these processes lead to a coherent end-to-end model and how this may address the basic question.     }


\graphicspath{{./IMAGES/}{./SOURCE_FILES/IMAGES/}}

\section{Introduction}

The remarkable pace of the discovery and characterization of exoplanets over the last 20 years suggests that a comprehensive, empirically verifiable theory of planet formation may now be possible.    Planet formation is a complex process involving a series of quite distinct pieces of physics and chemistry on physical scales ranging from micrometers to hundreds of AU.  As the other chapters in this section clearly show, each of these links in the long chain leading from planet formation to their observed dynamical, structural, and chemical properties require theoretical solutions to a number of deep problems.  While the connection between how planets form and their ultimate physical properties and chemical composition is at present, poorly understood, rapid progress is now being made. There are several general reasons for optimism.  

On the observational side,  the discovery of over 3000 exoplanets with thousands more to come  has revolutionized 
our understanding of planet formation and properties \citep{Mayor1995,Queloz2000,Pepe2004,UdrySantos2007,Howard2010,Howard2012,Batalha2014,Bowler2016}.
Statistical samples are now large enough that the properties of at least 4 planetary populations (hot and warm Jupiters, mini-Neptunes, and super Earths) can  be discerned.   We are also starting to link the properties of stars with their planetary systems. There are, compliments of the ALMA revolution, major advances in high resolution and chemical studies of protostellar disks in which planets form.   The chemical and physical properties of the outer regions of disks are  being probed for a wide range of host stars, and this has already yielded the surprising fact that either low dust/gas
ratios or a very large fraction of carbon (a factor of 100) are missing from the gas phase.   JWST will tackle the inner regions of disks, as well as the atmospheres of exoplanets. In the solar system, the Juno mission has for the first time, revealed the existence of a core within the planet which may be more dilute than expected. Thus, the physical and chemical processes leading to planet formation as well as the resulting populations  can now begin to be studied and tested using a wide variety of ground and space based observatories and probe.   

On the theory side,  major advances over the last decade include the development of 
sophisticated theories of planetary migration, dust evolution and the growth of pebbles and planetesimals, 
 a deeper understanding of radiative heating processes, and the rise of astrochemistry as a tool to 
 probe the process of disk evolution and planet formation.  All of these advances have been made with the
 help of a growing arsenal of powerful and sophisticated computer codes.   
 A successful comprehensive theory now requires that it addresses an array of 
ever more stringent inputs and constraints.  

This progress has given rise to a series of important questions.   The key question that motivates all aspects of this
review is this.   Is it possible that despite this plethora of complex processes, is there still a clear thread 
that connects their composition and other physical properties with their formation?   Or have these links been 
erased as one process takes over from the previous one?   If there is such a connection, do chemical abundance
patterns of gaseous atmospheres - say the ratio of carbon to oxygen (C/O) abundance - reflect on formation conditions,
such as planet formation at ice lines?  Are the observed inhomogeneities in disks such as gaps and 
rings a consequence of planet formation, opacity transitions, other?  
The transport of angular momentum is central to disk evolution and planet formation 
so are there imprints of these mechanim(s) left on planetary populations?  
Given that information about planetary compositions is most likely to come 
from observations of their atmospheres, to what degree are the bulk characteristics of the interiors of planets linked
to the composition of their atmospheres?  These are just a few of the interesting questions that arise in 
understanding this story.   

The goal of this review is to outline progress in the connections between planet formation in evolving (dynamically and chemically) disks
and the physical and chemical properties of the end product.  Most of the analysis addresses processes that occur while planets are forming in their natal disks.  We will step along from  the basic properties of evolving protostellar disks and planet migration and formation and end up with predictions about populations of planets whose statistical properties
that can be confronted with the data. This  section of Springer's  {\it Handbook of Exoplanets} contains excellent reviews of various aspects of planet formation, with an overview by Armitage (see also \cite{Armitage2010}). In addition, the reader may also consult a number of recent review articles on the various pieces of this problems including \cite{Testi2014} for disks and dust evolution; \cite{Turner2014} for disks and angular momentum transport; \cite{KleyNelson2012} for planet migration, and \cite{Raymond2014}, \cite{Helled2014}, and \cite{Benz2014} for planet formation.  

\section{Observational Constraints}

The basic properties of exoplanets can be conveniently summarized in just 4 or 5 important diagrams.  
The first, and perhaps most fundamental is the mass- semimajor axis (M-a) diagram which can
be conveniently divided  into three or four planetary populations \citep{ChiangLaughlin2013,HP13}.  The fact that Jovian planets pile up at a characteristic orbital radius of 1 AU, with slightly smaller mass hot Jupiters inside 0.1 AU is good evidence that these massive planets must have moved substantially during their formation in disks (see
chapter by Izidoro and Raymond). The theory of planet migration that has arisen to explain this rests on ideas of how planet-gas gravitational interaction and disk angular momentum transport works. 
The recent discovery of a Hot Jupiter with an estimated mass of $ 1.66 M_{Jup}$  orbiting a young, weak-lined, T-Tauri star Tap 26 - a system that 
is only 17 Myr \citep{Yu2017}  -  at 0.0968 AU has been interpreted as evidence for Type II migration of the planet while in its host disk.  This
is too little time for planet-planet scattering processes to have taken place.  

A second breakthrough are the mass-radius (M-R diagram) relations governing planetary structure that are now being uncovered so that planetary structure and composition can, for the first time, be explored \citep{Howard2013,Rogers2014,ChenKipping2017}.   An important issue here is that it is the composition of the materials accreted onto forming planets, in particular the overall elemental abundances, plays a major role in determining the radius of a planet for a given mass. This is especially true for low-mass planets, whose radii depend sensitively on whether the planets are rocky, have substantial water contents, or have retained atmospheres. Knowledge of planetary composition will soon be greatly enhanced as JWST and other observatories make precise measurements of the composition of planetary atmospheres.

The third major diagram is the so-called  planet-metallicity relation \citep{FischerValenti2005,WangFischer2015} which says that massive planets are more likely to be detected around stars only if they have sufficiently high metallicity (solar and above).  These authors found that for a limited range of stellar masses (0.7 - 1.2 M$_{\odot}$) that the probability of a star to host a giant planet scaled as the square of the number
of iron atoms;  $ P_{planet} \propto N_{Fe}^2$.  Later studies, carried out for a wider range of stellar 
masses, found that more massive stars also tend to host Jovian planets, with the scaling $P_{planet} \propto N_{Fe}^{1.2 \pm 0.2}  M^{1.0 \pm 0.5} $ 
\citep{Johnson2010}. The most recent research affirms a strong planet-metallicity relation for Jovian planets while stars of all masses
and metallicities host low mass planets.  These findings suggest that low mass planets can form in all disks but that only a fraction of these in 
high metallicity, or in sufficiently massive disks can grow into massive planets within the disk lifetime  \citep{IdaLin2005,HP14}. 

The fourth major diagram is the eccentricity distribution of planets which shows that large eccentricities accrue to a a significant number of massive exoplanets. The median value of this eccentricity is very high $ \simeq 0.25 $.  The eccentricity of single massive planets can be attributed to planet-planet  scattering interactions after the gas disk has been dispersed  \citep{Chatterjee2008,JuricTremaine2008}.   

Another important result is the observed misalignment between the orbital plane of a traversing planet and the equatorial plane of the rotating star measured via the Rossiter-McLaughlin effect (see chapter by Tibaud). Roughly 1/3 of hot Jupiters show such misalignments. This raises an important question: Did these planets arise through dynamical interactions after migration in the disk had placed them in close-in orbits? Or did they arrive at these innermost orbits by some dynamical process such as the Kozai mechanism coupled with tidal friction?

In the latter case, a distant companion star can cause eccentric motions of a planet whose orbit can shrink and circularize drastically with time due to 
tidal interaction with the star, leading to close-in Jupiters with high eccentricity \citep{FabryckyTremaine2007}.  The observation of a hot Jupiter in orbit around a young T-Tauri star mentioned earlier suggests that at least in some cases, migration in disks can quickly move massive planets into close in orbits, although whether these would be perturbed out of plane would depend on subsequent planet-planet interactions.   It may be that the elemental abundances of such planets will ultimately discriminate between planets brought in via disk processes, sampling materials from the inner disk regions, as compared to scattered bodies originally formed in outer disk regions whose compositions reflect the dominance of ices.   

Finally, one of the most prominent dynamical features in the M-a diagram are the numerous, extremely compact systems that are well aligned and having 
short periods \citep{RangMargot2012,HansenMurray2013,ChiangLaughlin2013}.  Although the spacings between orbital pairs seems
to be random, nevertheless, there is an abundance of them that are just wide of major mean motion resonances (MMRs) and a lack of such pairs
just inside these \citep{Lissauer2011,Fabrycky2014}.  One of the explanations for this behaviour is the effect of planet-planetesimal disk interactions
on trapped, resonant pairs of planets (e.g. 2:1) \citep{ChatterjeeFord2015}.  It is clear therefore, that the M-a diagram is a composite recording both
the history of planet-disk 
evolution as well as planet-planet and other dynamical interactions.  

\begin{figure*}
\centering
\includegraphics[width=\textwidth]{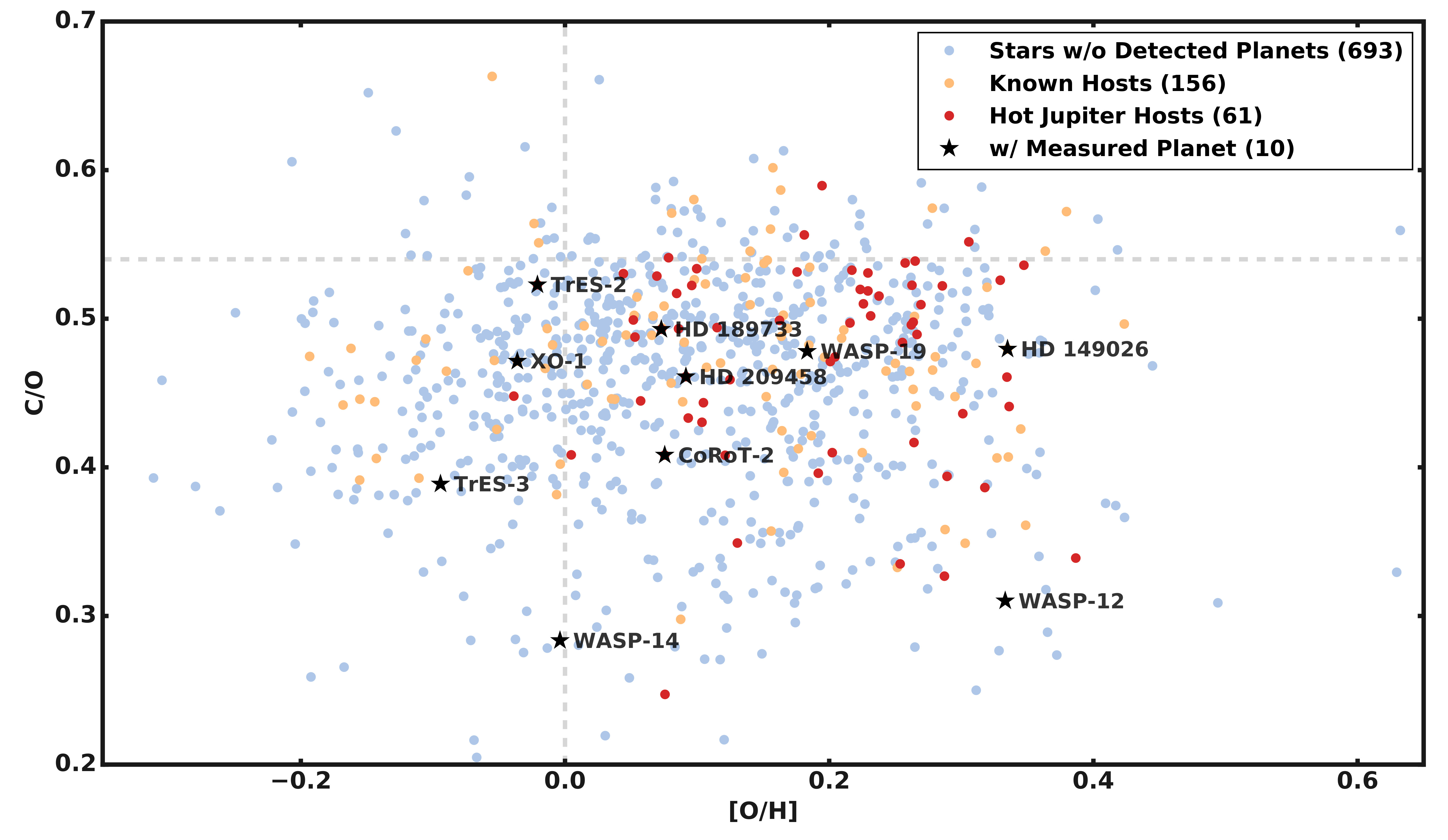} 
\caption{The distribution of C/O and O/H for a wide range of stars with and without planets. The dotted lines denote the solar values (C/O $=0.54$). Clearly there is no preference in C/O or O/H in the formation of planets, and particularly for hot Jupiters. Figure from \citet{Brewer16}, AJ, 153, 83. Reproduced with permission \textcopyright AAS.}
\label{fig:01}
\end{figure*} 

Turning now to the host stars of planetary systems, their chemical composition and radiation fields are essential external inputs for models.  Observations of the metallicity distributions and element ratios such as C/O and C/N of stellar atmospheres \citep{Brewer16} inform us about the distribution of element abundances in the initial accretion disks out of which both the star and its retinue of planets formed.  These materials were accreted onto the planet as it migrated through the disk.    Figure \ref{fig:01} shows the C/O and O/H ratios of 693 stars associated with detected planets (eg. Hot Jupiters), indicating that our Sun's C/O ratio is high compared to many planet bearing stars.  The figure therefore provides  information about the range of compositions and metallicities of the initial disks that hosted their forming planets. The difference between the composition of Jovian planets, and the host star metallicity, is most readily understood as a consequence of where and how planets accreted most of their  gas - indicating the possible role of ice lines as places where planets acquired most of their gas (\cite{Oberg11}, \cite{Madu2014}, Bergin and Cleeves review). The X-ray luminosities of protostars control the ionization state of their protostellar disks.  This is another key external stellar control parameter  for planet formation in that the ionization drives disk chemistry, which controls the extent of so-called dead zones in disks (regions free of turbulence driven by magnetic instabilities). The UV irradiation of stars also controls the disk lifetimes due to photoevaporation processes \citep{Gorti2016}.
 
Surveys will increasingly inform us about the distribution of protosellar disk masses and their lifetimes \citep{Haisch2001,Hernandez2007, Hartmann2008, Andrews2010}.  The distribution of disk masses is related to the initial conditions for disk formation and arise from the range of dense core masses and their level of magnetic braking and internal turbulence that will shape their collapse into protostellar disks (eg. \cite{Seifried2015}, review by \cite{Li2014}).   
The distribution of disk lifetimes is related to a combination of the processes that carry of disk angular momentum (turbulence, disk winds, or spiral waves)  as well as by disk photoevaporation 
processes that will ultimately dissipate them.   
There is growing understanding of how dust evolves in disks and of the changes in chemical composition as a function of disk radius,  arising from the appearance of various opacity transitions and ice-lines  (see Chapters by Bergin and Cleeves, Andrews and Birnstiel).  

\begin{figure*}
\centering
\subfloat[High resolution imaging of the HL Tau disk at mm-wavelengths. The resolution afforded by ALMA has given us an impressive look at the structure of protoplanetary disks. Figure from \citet{ALMA2015}, ApJ, 808, L3. Reproduced with permission \textcopyright AAS.]{
\includegraphics[width=0.5\textwidth]{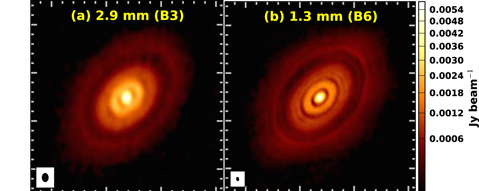}
\label{fig05a}
}
\subfloat[Detection of the CO ice line in the TW Hya disk. The position of the CO ice line is indicated by the inner gap of the N$_2$H$^+$ emission because it can be easily destroyed by gaseous CO. Figure from \citet{Qi2013b}, Science, 341, 630. Reproduced with permission \textcopyright AAAS.]{
\includegraphics[width=0.5\textwidth]{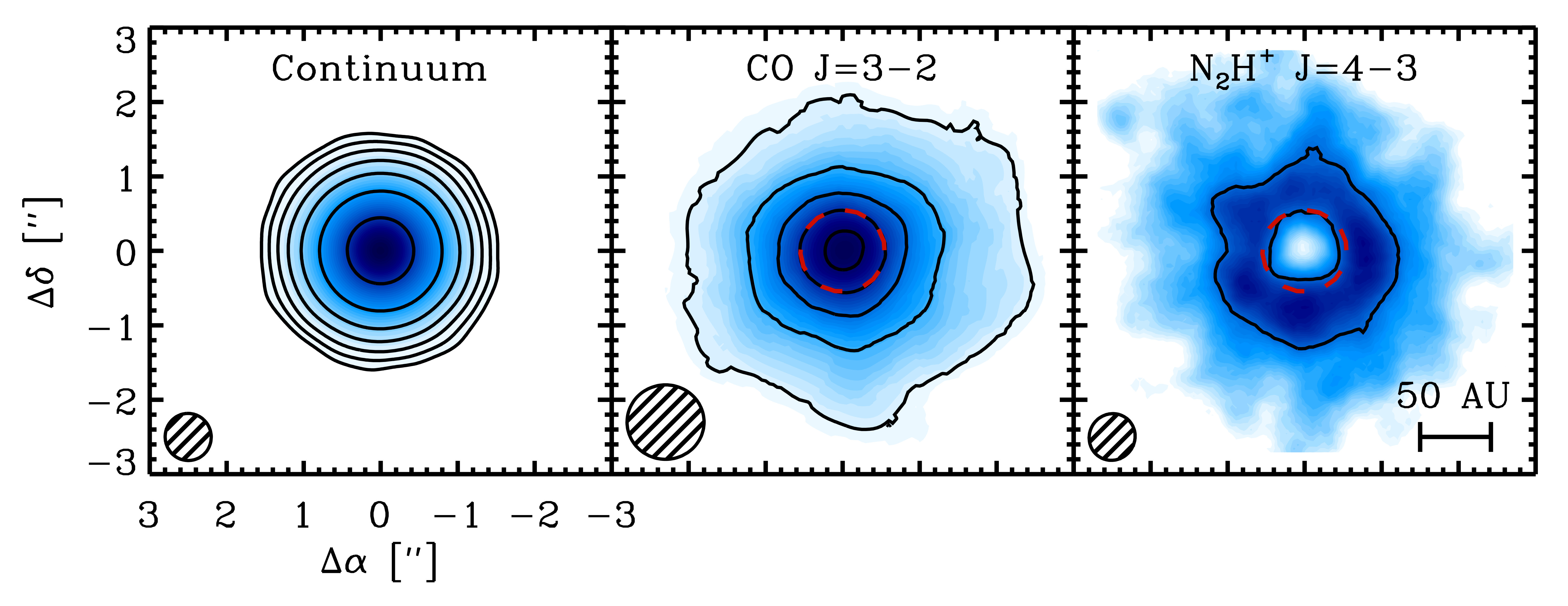}
\label{fig05b}
}
\caption{Examples of observational studies of the physical and chemical structure of protoplanetary disks.}
\label{fig:05}
\end{figure*}

One of the great observational surprises from ALMA is that disks have turned out to be far from the smoothly varying structures pictured in highly idealized theoretical models for accretion disks for decades.   ALMA observations as an example, show that disks host a  large number of symmetric ring and gap structures, as well asymmetric structure such as spiral waves and lopsided dust distributions revealing that density and temperature inhomogeneities dominate.  It is not yet clear whether these structures are the consequence of disk physics, or the result of planet formation.   Dust also has significant radial drift with respect to the gas in disks.   Figure \ref{fig05a} shows the now famous image of HL Tau with its series of either 5 \citep{Tamayo2015}, or 3 \citep{Zhang2015} gaps, whose origin has a variety of possible explanations ranging from the appearance of various ice lines (opacity transitions) to the perturbing influence of planets that are carving out gaps or creating pressure bumps into which dust gathers.  In Figure \ref{fig05b}, we see observational evidence for the existence of the CO ice line in TW Hya.   These inhomogeneities have important implications for planet formation in that they can give rise to dynamical traps for migrating low mass planets, as well as traps for rapidly moving dust.   
Finally, observations of debris disks are telling us about the degree to which carbon was frozen out and stored in planetesimals. These can retain imprints of planet formation and disk chemistry processes (see chapter by Wyatt).  

In short, there is now a wealth of statistical data on properties of stars, exoplanets, and protoplanetary disks that can be brought to bear on constructing a comprehensive picture of planet formation.

\section{Physical and chemical components of an end-to-end model}

The pioneering steps  towards connecting the data in M-a diagram with a statistical treatment
 for planet formation  in a core accretion model were taken in the first population synthesis paper of   \cite{IdaLin2004}. The intent of this approach was to model the evolution of planets in the M-a diagram using planet formation theory coupled to a statistical treatment of the initial conditions - the primary one being the distribution of disk masses.  Differences between predicted and observed populations then offer insight into how theory needs to be further developed  \citep{Benz2014}.   This is perhaps the most important first way that theory could be tested given that; (i)  only the initial conditions (eg. disks distributions) and final results (planetary systems in M-a) diagram are directly observed and that planet formation is not (yet), and (ii) the diversity of planetary properties arises in part from distributions of initial controlling parameters (eg. disk masses, metallicities, ionization rates, etc.).  This work  was followed up for stars of different metallicity and masses \citep{IdaLin2004b,IdaLin2005}.  These were then improved by more comprehensive treatments of various migration processes including an analysis of what is needed to slow rapid migration \citep{IdaLin2008},  and an examination of the ability of the ice line to act as a potential migration trap \citep{IdaLin2008b}. Investigations of the effects of stellar masses on planet populations were carried out by \citet{Alibert2011}.     

The basic components of an end-to-end theory of planet formation that also includes the chemical composition of newly formed can be briefly summarized.

-  (i) Adopt a model for the structure and evolution of protostellar disks, from the initial conditions (reflecting their formation), through disk evolution due to the proposed mechanism of angular momentum transport, to the end phases in which photo evaporation of disks leads to their rather quick demise 3 - 10 Myr after their formation.  The majority of treatments of angular momentum transport in disks assume that disks are turbulent, and that therefore it is ``turbulent viscosity'' that transports angular momentum \citep{SS1973,LBP1974}, the source of the turbulence being the magneto-rotational instability (MRI)  \citep{BH1991}. It has long been known however, that for ideal MHD, magneto-centrifugal disk winds can be more effective than even turbulence in transporting away disk angular momentum \citep{BlandfordPayne1982,PudritzNorman1986,PelletierPudritz1992,Pudritz2007}.   Recent breakthroughs in non-ideal MHD effects in disks show that MRI turbulence may be entirely suppressed in the central, planet forming zones of disks  $\le $ 10 AU) leaving only an MHD disk wind to transport out the angular momentum \citep{BaiStone2013,Gressel2015}.   The wind picture of angular momentum transport may have profound consequences for planet formation and migration (see Nelson's chapter).  Finally, disks are likely to be self-gravitating in their early stages of formation, and therefore spiral waves will appear which can be highly effective in transporting disk angular momentum radially \citep{Li2014}.

-  (ii) Prescribe the evolution of solids within such disks.   Dust may arrive in the disk during 
disk formation by the collapse of an initial  protostellar core, or form as the result of a condensation sequence wherein minerals appear at different disk radii depending on their condensation temperatures.   Subsequent dust settling into the midplane leads to rapid coagulation. Dust grains will grow due to agglomeration near the disk midplane, while at the same time undergoing radial drift due to drag forces. Radial drift changes the dust to gas ratio throughout the disk and will help dictate where planets may form (see chapter by Andrews and Birnstiel).  

- (iii) All planets, whether terrestrial, or massive, start by the accretion of solids into smaller mass embryos and cores.  The nature of solid accretion could reflect either collisions of planetesimals to build oligarchs, pebble accretion onto rapidly growing bodies, or a combination of these. The composition of these materials will play a basic role in determining the M-R relation.

- (iv) Migration of embryos and forming planets. Theories of migration for bodies with masses that are too small to open gaps (Type I migration) focus on two kinds of torques due to planet-disk interaction:  wave torques due to Linblad resonances at some distance (a few Hill radii) from the forming planet (usually resulting in inward migration) and co-rotation torques exerted by gas orbiting very close to the forming planet, generally resulting in outward migration. Real disks also have inhomogeneities in temperature and densities, and these prove to be crucial in providing zones of "zero net torque" or planet traps.  

- (v)  Gas accretion onto massive cores leading to accretion runaways that quickly build Jovian planets.  The composition of gas accreted during this phase will determine a great deal about the properties of the Jovian atmospheres (see chapter by D'Angelo and Lissauer).   The latter is best followed using time dependent gas chemistry codes \citep{Fogel2011,Helling2014,Crid16a,Eistrup2016}. 

- (vi)  Gap opening, Type II migration , and the end of accretion from the disk.  The late accretion from planetesimals may affect the chemical composition of the atmospheres.  

- (vii)  End of planet- gas disk interaction that arises from the photo evaporation of the disk.  This does not yet mark end of planetary chemical enrichment of atmospheres. 

- (viii) The dynamical evolution of the gas-free planetary system in which planet-planet interactions will rapidly lead to high eccentricities, the loss of mean motion resonances, and probably the loss of some planets (chapter by Morbidelli). The scattering of planetesimals onto colliding trajectories with gas giants may lead to considerable metal enrichment in Jovian atmospheres.  

- (ix) The structure of a planetary atmosphere depends on its pressure-temperature (P-T) profile as well as its chemical composition.  For massive planets, this is usually computed using chemical equilibrium models based on the elemental abundances of gas and solid materials delivered to the forming atmosphere during its formation (eg. see Madhusudhan chapter).  For Super Earths, the composition of secondary atmosphere that forms due to outgassing will depend on the accretion history as well, as volatiles undergo outgassing from the newly formed planet.

\section{Disk formation and initial chemical composition}

Disks with radii $ > $ 30 AU have now been observed during the earliest phases of protostellar evolution - the Class 0 and Class I sources.  (eg. \cite{Tobin2015} - class 0; Andrews and Birnstiel chapter).  Given that the star forming cores within filamentary molecular clouds have a wide range of masses and angular momenta, one expects a wide range in protostellar disk properties.

Recently,  \cite{Bate2012} published the highest resolution (down to the opacity limit for fragmentation - a few Jupiter masses), radiation hydrodynamics simulation of a forming star cluster.  The initial low
mass, cluster forming clump (500 $M_{\odot}$ ) had an initial radius of 0.40 pc and temperature of $10^o$ K.  The resulting initial mass function of the stars closely followed the observations \citep{Chabrier2005}.   A comprehensive study of the properties of disks formed in this simulation have now also been published \citep{Bate2018}.   The disks show an enormous diversity in types and sizes. Systems can be formed by a wide range of  processes including filament fragmentation, disk fragmentation, dynamical processing, accretion and ram pressure stripping.  Disk morphologies include warped and eccentric disks.   Disk masses increase until $10^4 $ yrs with masses up to 0.5 $M_{\odot}$.   Disk masses range from $M_d/ M_{\odot} ~ 0.1 - 2$ for times $\le 10^4$ yrs,  after which they decline.  Thus, disk masses at these early times are some 30 - 300 times more massive than they are during the Class II phase (when the are $\sim$1 Myr old). The surface density profiles are flatter than the Minimum Mass Solar Nebula, with $\Sigma_d \propto r^{-1}$,  rather than  the classic MMSN $r^{-3/2}$ 

\begin{figure*}
\centering
\subfloat[Cumulative mass distribution of protoplanetary disks. The disks produced in the numerical simulations of \citet{Bate2018} (black line) tend to be more massive than the observed systems because they are much younger than the observed systems. Figure from \citet{Bate2018}, MNRAS, 475, 5618. Reproduced with permission \textcopyright Oxford University Press.]{
\includegraphics[width=0.5\textwidth]{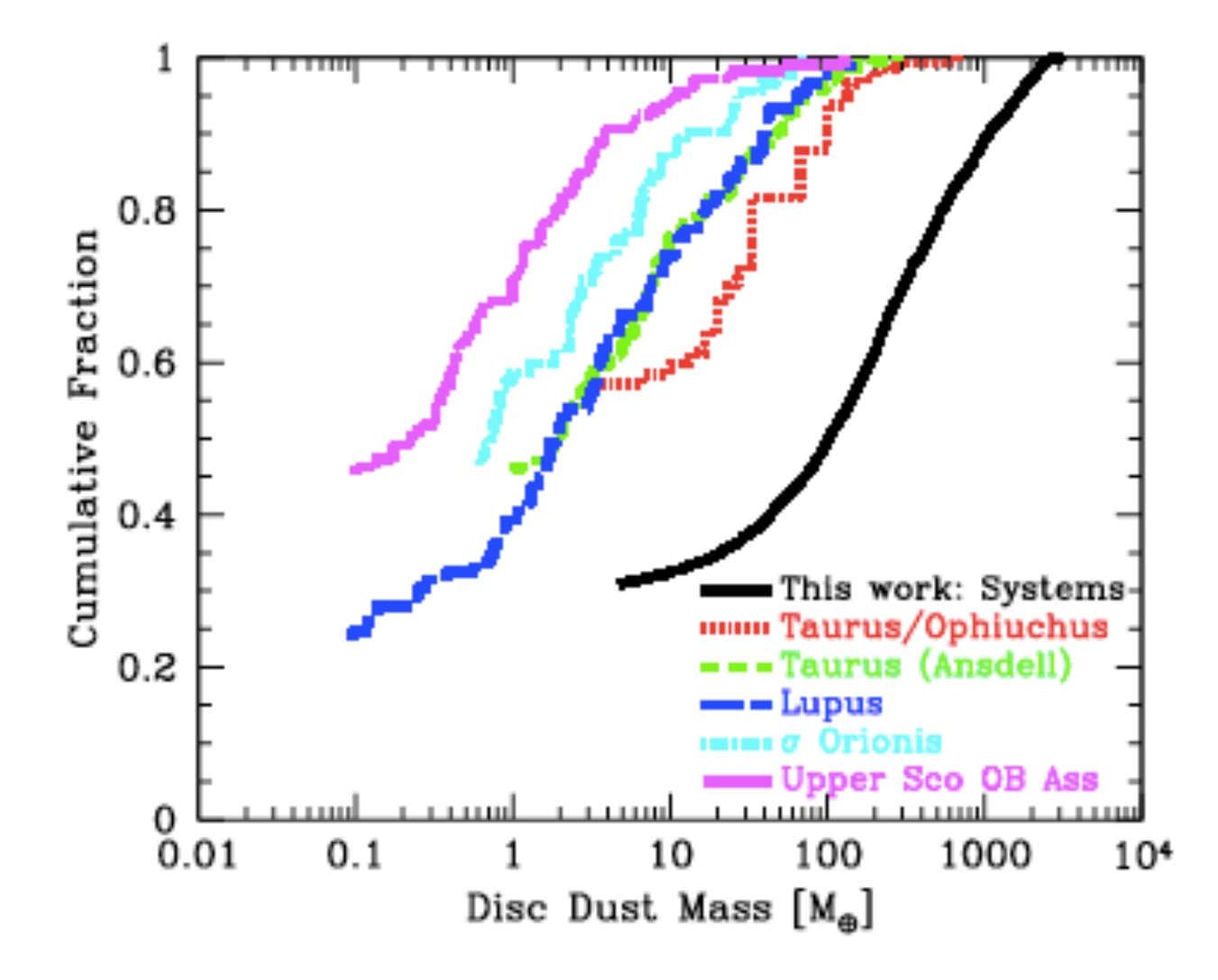}
\label{fig:DiskForm01a}
}
\subfloat[Three-dimension snapshot from an MHD collapse calculation for the collapse of a 2.6 solar mass core. Black lines are magnetic field lines, blue coloration is of dense filaments bringing gas into the disk forming in the central region. The scale of the box is 1300 AU. Figure reproduced from \citet{Seifried2015}, MNRAS, 446, 2776.]{
\includegraphics[width=0.5\textwidth]{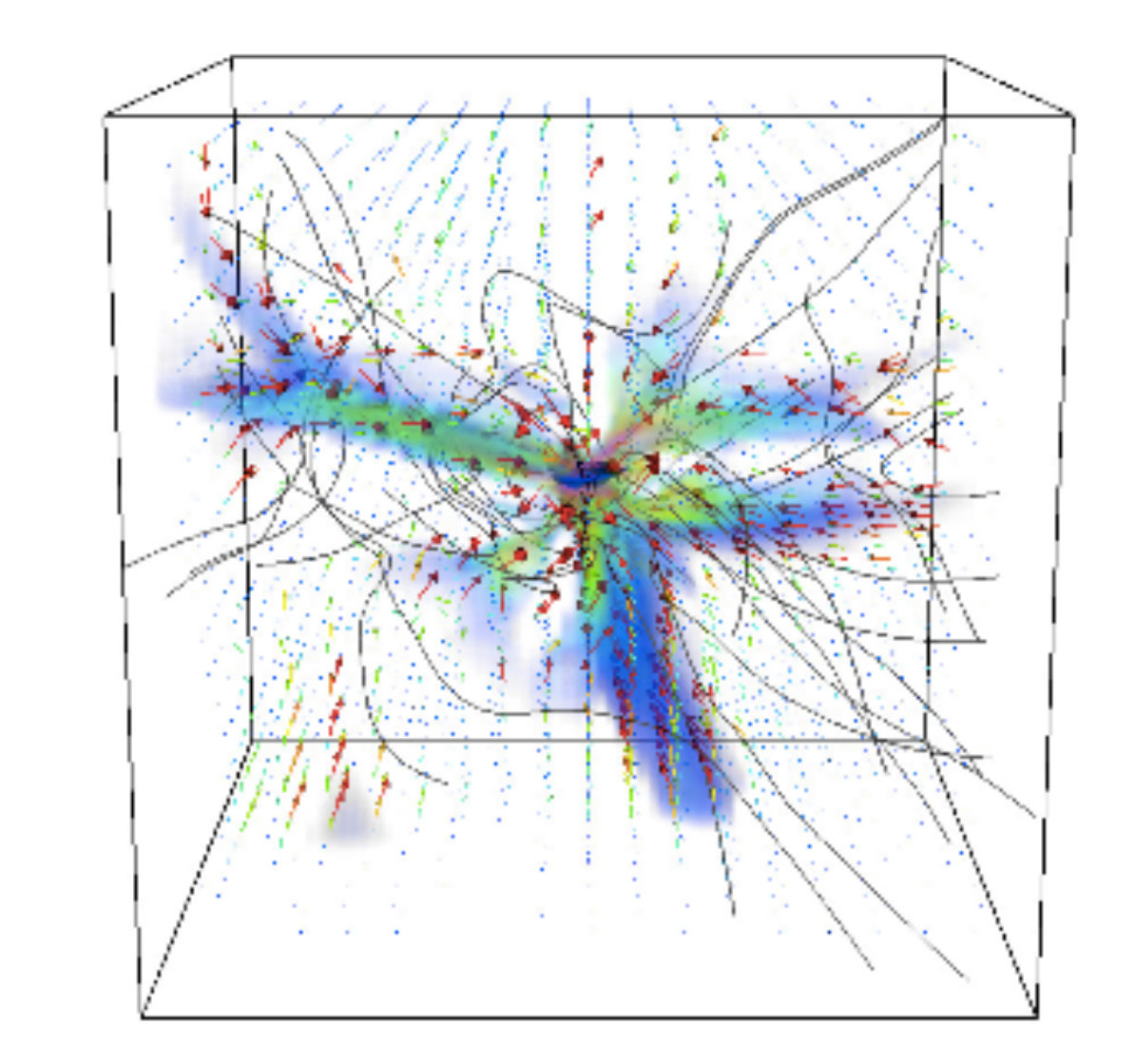}
\label{fig:DiskForm01b}
}
\caption{Left: Cumulative mass distribution for disks in radiative hydrodynamics simulations.  Right: Image of a disk forming in 
a turbulent MHD simulation}
\label{fig:DiskForm01}
\end{figure*}

These radiation hydrodynamics simulations did not include magnetic fields.  A reasonably strong magnetic field will strongly brake smooth, rotating clouds so that only very small disks can form - a result known as the ``magnetic braking problem''.  This can be resolved in turbulent simulations in which the magnetic torques are much reduced, leading to larger disks more resembling the hydrodynamic results \citep{Li2014,Seifried2015}.

The left panel of Figure \ref{fig:DiskForm01} shows the cumulative distribution of disk masses in Bate's (2018)  turbulent radiative hydrodynamics simulation (figure \ref{fig:DiskForm01a}). 
The distribution is roughly an order of magnitude more massive than observed disks  in various clouds - seen at much later times.  It is not clear how to connect the 
time of the simulation with the time in the observed clouds, but the latter pertains to objects much older than the newly collapsed disks in the simulation.
The right panel of the figure shows an image of a forming disk in a turbulent MHD simulation  \citep{Seifried2015} (figure \ref{fig:DiskForm01b}). 
Here too one sees a highly filamentary collapse process in which 5 or so filaments bring material to the forming disk while MHD torques are inefficient
in providing a magnetic brake at this earliest phase.  
More generally, collapsing, magnetized cores will launch magnetically driven outflows and winds as the disks are forming, and long before the final process of stellar assembly is complete \citep{BanerjeePudritz2006,Li2014}.   Thus, even  in the earliest stages, MHD disk winds will play an important role in the angular momentum evolution of these systems, and this can lead to profound effects on planet formation.

The dust and chemical composition of the prorotostellar core can, to some degree, be inherited by the disk.  Thus, whereas the largest part of dust growth will occur at the disk mid plane because coagulation is more rapid in high density environments, coagulation helped by ice coated mantles (eg. \cite{Ormel2009}, grows grains to several microns at core densities of $10^5$ within $\sim$1 Myr.  Dust can grow up to $\sim$mm sizes within the infalling envelopes (e.g. \cite{Jorgensen2009}).  

Chemical processing also occurs within the dense gas of star forming cores.   
In prestellar cores, the most abundant phase for molecules with elements heavier than hydrogen and helium is a solid.  It has been known for over 40 years \citep{GillettForrest1973} that infrared absorption of interstellar ices gives us a glimpse into the chemical composition of star forming material.  Water and CO ice were the first to be discovered, and represent the most abundant molecules after H$_2$. They are followed closely in abundance by CO$_2$ which was not found until the launch of IRAS because of strong absorption in the atmosphere \citep{Oberg11}. With newer space-based studies by ISO and Spitzer, larger, more complex hydrocarbons have been inferred in the infrared absorption of ices towards star forming regions \citep{Oberg2011b}. The formation of these hydrocarbons through the hydrogenation of frozen CO has been studied both theoretically \citep{Walsh2014,Vasyunin2017}, and in laboratory experiments \citep{Butscher2015,Chuang2018}, and represents the first steps towards pre-biotic chemistry. With the latest generation of telescopes these pre-biotic molecules have begun to be found around both young stars \citep{Jorgensen2012}, and in prestellar cores \citep{Ligterink2017,Rivilla2017}.

Whether these species survive to the protoplanetary disk is still debated. There are two primary pictures for the delivery of element from the prestellar core to the disk: `inheritance' and `reset'. In the inheritance scenario all molecular species that were formed in the prestellar core are delivered to the disk intact, while in the reset scenario there is a thermal event that breaks all molecules down to their base elements \citep{Pontoppidan2014}. 
As an example, detailed chemical studies of over 39 different molecules, grouped into 4 families of related molecules, have been carried out in the well studied, pre-stellar core  L1544, indicate that significant differentiation of C and N bearing molecules occurs.  Such studies holds great promise for understanding the initial chemical conditions before disks formed  \citep{Spezzano2017}.
Deuterated water could be a good tracer of these different process because its enrichment is favoured in cold, ionized environments like prestellar cores (eg. L1544) , and in a protoplanetary disk at large radii \citep{Cleeves2014}.  In the inheritance scenario the deuteration of water would be homogeneous across the disk, while in the reset scenario there would be a deuterium gradient. Of course the true answer may be somewhere between these two extremes.  Carbon deficiency in the solids throughout the solar system could be evidence of reset in the inner solar system, and inheritance in the outer solar system. The number of carbon atoms relative to silicon on the Earth is under abundant  by four orders of magnitude relative the ISM while comets like Halley are not similarly underabundant \citep{Bergin2015}. This could be evidence of thermal processing of material because while carbon generally exists in the solid phase in prestellar cores \citep{Bergin2015} if the grains are destroyed upon reaching the disk, the carbon would not re-condense as a solid in the inner solar system \citep{Pignatale2011}.

\section{ Disks:  structure, evolution, and chemistry } 

As the inflow of gas onto the forming disk comes to an end, the earliest stage of  disk evolution has the disk mass being comparable  to that of the protostar \citep{Seifried2015,Klassen2016,Bate2018}. The self gravity of the disk,  measured by the Toomre Q parameter ($ Q = c_s \Omega / 4 \pi G \Sigma $ where $c_s$ is the sound speed, $ \Omega $ is the local angular velocity of the disk, and $\Sigma $ is its surface density), is significant  ($ Q \simeq 1$  \citep{Kratter2008}).  In the first $10^5 $ yrs, the system evolves
from the Class 0  to the Class I state in which outflows are the most powerful.  Angular momentum transport in the disk arises from  spiral waves (due to the  gravitational instability of the disk launched when $ Q \le 1$), together with disk winds and turbulence.  It is during this time that the first stages of planetesimal formation would have already taken place.    

The subsequent phase of disk evolution in the much better studied Class II systems, involves thin, Keplerian disks that are in vertical hydrostatic balance.   The vertically averaged angular momentum equation that governs a disk undergoing a total stress $\bf \sigma $  is  \citep{PudritzNorman1986,Turner2014,Bai2016}, \begin{equation}
 \dot M_a {d \over dr} (r u_{\phi}) = {d \over dr} ( 2 \pi r^2 < \sigma_{r,\phi} > )+ 2 \pi r^2 \sigma_{z,\phi} \vert_{-h}^{+h} 
\end{equation} 
\noindent where the accretion rate is $ \dot M_a = 2 \pi r  \Sigma v_r$ for a radial inflow speed of the gas $v_r$, and the angle brackets in the first term indicate taking the vertical average of the torque by integrating over z.  The total stress has contributions from both turbulence, and the Maxwell stress of threading magnetic fields.  The first term on the right hand side denotes angular momentum flow in the radial direction, while the second term is angular momentum flow out in the vertical direction due to wind torques.  In the case of shear turbulence, the stress is  the average of the turbulent fluctuations, 
$ \sigma_{ r, \phi} =  -  \rho  \delta  v_r  \delta v_{\phi}   $.
In the presence of  a toroidal magnetic field $B_{\phi}$ in the disk, a radial field $B_r$ can also contribute to flow in the radial direction through  the Maxwell stress component; 
$\sigma_{r, \phi} =B_r B_{\phi}  $.   This possibility arises in recent models of non-ideal MHD wherein the Hall effect can produce an instability leading to a radial field component  (eg. \citep{Bai2014,Lesur2014,McNally2017}). A threading vertical component of the field $B_z$  however, exerts a torque on the disk with $\sigma_{z, \phi} = B_z  B_{\phi}  $ leading to
 an MHD disk wind, which is central to the action of the ubiquitous jets and outflows that accompany the formation of all young stars, regardless of their mass \citep{Frank2014,Ray2007,Pudritz2007}.  

 Physical models of accretion disks have focused heavily on the assumption that angular momentum is transported  by turbulent viscosity, first addressed in the seminal papers by \cite{SS1973}, \cite{LBP1974}.
 Here, the turbulent  arises from the shearing Keplerian flow and takes the form $ \sigma_{r, \phi} = \nu \Sigma r d \Omega / dr $. The effective viscosity of the disk $\nu$ can then be shown to scale with the disk scale height as $ \nu = \alpha c_s h$ with the famous $ \alpha $ parameter.    
Steady state disks then have a radial accretion rate $ \dot M_a $ , which, away from the inner boundary of the disk can be written as

 \begin{equation}
 \dot M_a = 3 \pi \nu \Sigma = const 
 \end{equation}

In order to drive an accretion flow at the rate observed to fall onto T-Tauris stars, $\alpha_{SS} \simeq 10^{-2} - 10^{-3} $.  The angular momentum is carried out radially leading to the slow, outward radial spreading of the disk from its initial state.  The energy that is available to drive the turbulence is given by the gravitational potential energy release across each annulus of the disk, which is dissipated as heat and radiated away.  Assuming that each annulus of the disk radiates as a black body then one readily derives that viscous heating results in an effective temperature of the disk  $\sigma T_{eft}^4= (3/8 \pi) \dot M_a \Omega^2$ and thus the scaling: $T_{eft} (r) \propto \dot M_a^{1/4} r^{-3/4} $.    

 The second source of heating is the radiation field of the 
 central star.   A flaring disk will intercept flux from the star,  and will be absorbed by the dust and re-emitted at IR wavelengths in the disk's surface layer .  Assuming this is a black-body process, the temperature then has a shallower fall off with disk radius $T(r) \propto r^{-1/2} $  \citep{HartmannKenyon1987}.   This can be extended by 
 considering that only the surface layers of the disk are directly heated by the star while the deeper parts of the disk are heated by radiation re-emitted from it, which
 are solved in concert with 
 hydrostatic balance that produces a flaring disk .  The result is a  surface temperature that scales as 
 $T_{surf} \propto r^{-2/5} $ while for the interior $T \propto r^{-3/7} $ for disk radii $ r \le 84 AU $ \citep{ChiangGoldreich1997}. Observations indeed show that the temperature distribution arising from viscosity are too steep to explain the mm and submm observations of disks, having an average temperature exponent $T \propto r^{-q} $, where
 for the dust,  $q_{dust} \simeq 0.5$  \citep{AndrewsWilliams2007}.  The temperature of the gas,  as determined by CO and [CII] line observations,  has a steeper radial decline with  $q_{gas} \simeq 0.85 $. Since the temperature profiles of dust and gas should be similar on the disk mid plane, the difference here suggests a decoupling of gas and dust at high scale heights above the disk \citep{Fedele2013}.   
 
 Unlike viscous heating, radiative heating from the central stars creates a hot surface layer on the disk atmosphere, and a much cooler midplane.  This has several important consequences for disk chemistry and dynamics in that the snow-lines for various species are 2-D surfaces that move outward in radius as one moves away from the disk midplane (see chapter by Bergin and Cleeves).  The disk radius at which the dominant heating mechanism of the disk transitions from viscous to radiative heating is called called the ``heat transition'' (eg. \cite{Lyra2010, HP11} ), which we will denote $r_{HT}$.  

Disks are not static structures and their time evolution due to the long term action of disk viscosity is well known. In general, this is determined by solving the continuity equation for the surface density of the disk, together with the disk angular momentum equation.  For disks driven by purely viscous torques, 
the equation describing the evolving surface density profile $\Sigma(r,t)$ of a protoplanetary disk becomes: 

\begin{equation}
 \frac{\partial \Sigma}{\partial t} = \frac{3}{r}\frac{\partial}{\partial r}\left[r^{1/2}\frac{\partial}{\partial r}\left(r^{1/2} \nu \Sigma\right) \right]\, .
\label{DiskEvolution} \end{equation} 
 
\noindent This diffusion equation describes accretion as the result of a diffuse process driven by the turbulence.  

Since the surface density evolves with time, the accretion rate $\dot M_a$ must also be affected, and in fact decreases with time.  The disk will also lose mass due to photoevaporative processes that are driven by X-ray and UV radiation from the star. The combined effects of accretion and photo evaporation can be combined in a single, time dependent equation for the evolution of the disk's accretion rate \citep{Pascucci2009,Owen2011};  

\begin{equation} 
\dot{M}(t) = \frac{\dot{M}_0}{(1 + t/\tau_{\rm{vis}})^{19/16}} \exp\left(-\frac{t - \tau_{\rm{int}}}{t_{\rm{LT}}}\right)\;,
\label{ViscousAccretion} 
\end{equation}
which includes a viscous evolution term multiplied by an exponentially-decreasing photoevaporation factor. In equation \ref{ViscousAccretion}, $\tau_{\rm{vis}}$ is the disk's viscous timescale, $\dot{M}_0$ is the accretion rate at the initial time $\tau_{\rm{int}} = 10^5$ years, and $t_{\rm{LT}}$ is the disk's lifetime \citep{APC16a}.  In this equation the contribution due to viscous diffusion arises from the analytical model by \cite{Chambers2009} for the evolution of a viscous, irradiated disk.  The exponential factor is due to  rapid photoevaporative truncation of the disk as modelled  by \cite{HP13} who showed that without a sharp cutoff of viscous evolution planets undergo too much migration and accretion to be able to match the distributions in the M-a diagram.  We note that other authors have used different expressions for cut-offs, such as a finite time cutoff to zero \citep{Ruden2004}.  

We see that the distribution of disk lifetimes $t_{LT}$ directly impacts the accretion histories of disks and stars through sharp photoevaporation driven cutoffs of the disk surface density. Thus planet-disk interaction ceases fairly quickly once photoevaporation sets in and planets cease their migration.  There is another crucial aspect of evolving radiatively heated disks.   Since the temperature of the inner viscously heated part of the disk must decline with time (since $T_{visc} \propto \dot M_a^{1/4}$ ), the heat transition radius $r_{HT}$ moves inwards with time as well.  The heat  transition radius turns out to also play the role of a planet trap - wherein fast moving low mass planets underling Type I migration are trapped at a point of zero-net torque.  We discuss this below.   
 The viscous evolution picture of disks requires an explanation of how turbulence can be
excited and maintained in disks.   Hydrodynamic Keplerian disks are highly stable to various kinds of perturbations but there is a large literature on how  turbulence could be excited.  
The central pillar on which most thinking about turbulence in disks rests has been the magneto-rotational instability (MRI).   In his magisterial treatment of instabilities in magnetized rotating fluids, Chandrasekhar ({\it Hydrodynamic and Hydromagnetic Stability}, 1960) notes a striking fact about the stability of so-called Couette flows (fluid flow between two rotating cylinders) .  For purely hydrodynamic systems, the
 well known Rayleigh criterion for fluid stability dictates that the specific angular momentum (i.e., angular momentum per unit mass, $j = v_{\phi}r = \Omega r^2 $ should increase with radius for stable hydrodynamic flows.  However, if one threads a rotating Couette flow with a magnetic field, this criterion is profoundly changed:  stability requires that $\Omega$ must be an increasing 
 function of radius - even in the limit of vanishing magnetic field strength.  The seminal paper by \cite{BH1991}, in working on the stability of magnetized accretion disks, rediscovered 
 this result.  In the astrophysical context, there is no system that we know of, with the exception of small boundary layer regions,  whose angular velocity decreases increases with radius (eg. galactic rotation curves $ \Omega \propto r^{-1}$, Keplerian disks $\Omega \propto r^{-3/2}$ ). Accretion disks, it follows, should be highly unstable to MRI.   Growth rates for the most unstable modes in a thin disk are $3/4 \Omega$ \citep{BH1991} with a vertical wavelength $\lambda_z = 2 \pi v_A / \Omega $  where $v_A = B_z / (4 \pi \rho )^{1/2} $ is the Alfven speed in the magnetized gas. If a toroidal field component is also present, then the field becomes buoyant for a plasma "beta" parameter (the ratio of thermal to magnetic pressure in the gas $ = 8 \pi \rho c_s^2 / B^2 $ ) for sufficiently strong fields $ \beta \le 10$ \citep{TerquemPapaloizou1996} 
 and rises out of the disk into the disk corona.  
     
 The high column densities of protoplanetary disks prevent much radiation from penetrating the disk, leaving it poorly ionized at the midplane.  Thus non-ideal MHD processes such as Ohmic losses, ambipolar diffusion, and the hitherto little investigated Hall effect, all take their toll on the coupling of magnetic fields to gas.   The region in the disk where these non-ideal effects conspire to reduce or eliminate the MRI instability is known as the dead zone \citep{Gammie1996}.  
 
 Disk evolution in these regions are essentially dead to MRI turbulence, and while very low level turbulence may still be excited by various hydrothermal instabilities \citep{Flock2012}, in the presence of a weak threading vertical field, angular momentum is driven primarily by a disk wind from a largely laminar disk \citep{BaiStone2013,Gressel2015}.  The wind is launched from a thin, highly ionized  region on the disk surface.   Radial flow is possible in such laminar disks if the Hall effect is radial, laminar transport of angular momentum occurs.   One of the most distinct aspects of the Hall effects is that the direction of transport of the magnetic flux in disks depends on the polarity of the threading poloidal field component $\bf {B_p} $ with respect to the disk rotation axis.  If its direction is parallel to { \bf $ \Omega$ }, then flux transport 
 is inwards, and if anti-aligned, outwards \citep{BaiStone2017}.  Since the flux distribution affects the strength of the wind torques,
 these Hall effects could be significant for the physics of Type I migration.    
 In all situations, it appears that disks do not support MRI turbulence out to distances of 10 AU for standard conditions.  This dead zone radius $r_{DZ}$ must evolve with time as the disk thins out. 

 \subsection{Disk ionization, turbulence, and angular momentum transport} 
 
The ionization of the disk by  stellar X-rays, external cosmic rays, and the decay of radionuclides mixed in with the gas plays a central role in the coupling of the magnetic field - and hence the genesis of MRI turbulence - to the disk.   Disk chemistry is also primarily driven by ionization processes (see Chapter by Bergin and Cleeves). Thus, disk chemistry and angular momentum transport are highly coupled, and as we will see, should therefore be connected to the ultimate element compositions of forming planets.  

Non-ideal MHD effects arise from the finite diffusivity of fields in the background gas.   Ionization fractions are highest at the disk surface and decrease with increasing optical depth as one penetrates down to the disk mid plane.  Thus UV and X-rays are absorbed at column densities of 0.01 and 10 g cm$^{-3}$ respectively.   The greatest penetration can be achieved by cosmic rays (CR) which are attenuated by column densities of 100 g cm$^{-3}$ \citep{UmebayashiNakano2009}.  Unlike X-rays however, CR can be scattered by MHD turbulence.  By decomposing MHD perturbations into their three basic modes (slow, Alfvenic, and fast),  it has recently been shown
that gyroresonance with the fast modes (sound waves compressing the magnetic field)
 is the dominant scattering process \citep{YanLazarian2002}.   
 This is expected to occur for CR propagation  through protostellar and disk winds, as is evidenced by the lack of CR driven chemistry in protostellar disks \citep{Cleeves2013}.  
 
 As one moves from the surface to ever greater densities approaching the disk mid plane, 
 first dust grains, then ions, and finally the electrons decouple from the magnetic field.   The degree of coupling is measured by three different
 magnetic diffusivities \citep{SalmeronWardle2003};  ambipolar diffusion in the surface
low density regions where ions and electrons are well coupled  ($\eta_A$),  the Hall effect at intermediate densities 
where the ions are decoupled from the fields through insufficient collisions with the neutrals ($\eta_ H$), and at the
greatest depths and densities Ohmic diffusion where even the electrons become decouples ($\eta_O$).  While both 
ambipolar and Ohmic effects behave like diffusive processes, the Hall effect is different in principle.  It drives the field
lines in the direction of the current density with a tendency to twist that can give rise to 
non diffusive dynamical processes, such as the generation of a toroidal field from a radial component. 

The diffusivities depend upon the ionization of the disk, and it is here that models of disk ionization driven chemistry
can play a key role.  As an example, the Ohmic diffusivity depends on the electron fraction $x_e$ and disk temperature as
$ \eta = 234 T^{1/2} / x_e$ cm$^2$ s$^{-1}$.   As one moves towards the disk mid plane, the Ohmic diffusivity grows as 
the X-rays are screened.  Similarly, as the disk evolves, the column density at any radius decreases with time, shifting
the region of Ohmic dominance inwards allowing turbulence to appear.   The temperature at the disk mid plane, where planetary materials are gathering,  is related to the effective temperature of the disk as $ T_{mid} = (3 \tau / 4)^{1/4}T_{eff}$ where $ \tau = \kappa_o \Sigma / 2$ is the optical depth and $\kappa_o$ is the disk's opacity.  Chemistry codes that can follow disk ionization with time are therefore essential (eg. \cite{Crid16a,Crid16b}).  

The damping of MRI instabilities can be measured by the ratio of the growth rates to the damping rates predicted by 
these diffusivities.  These are the so-called  Els$\ddot a$sser numbers for each effect: $ A_m =   v_A^2 / (\eta_A \Omega)$, 
$\Lambda_H=  v_A^2 / (\eta_H \Omega)$, and $\Lambda_O =  v_A^2 / (\eta \Omega)$.    Damping of the turbulence will occur if these numbers take a values of typically less unity (see \cite{Turner2014} for a review).   In the case of Ohmic diffusion,  this comparison of damping and growth rates can also be expressed in terms of a comparison of physical scales, 
namely, that the diffusion will erase fluctuations on a scale smaller than $\eta / v_A$, while the fastest growing mode in the disk has a wavelength of $ 2 \pi v_A / \Omega$.  
 
 The appearance of a dead zone in disks has important implications for planet formation and chemistry.   In order to 
 maintain a constant accretion rate throughout the radial structure of a disk at any time,  the 
 relative roles of turbulence and disk winds in transporting angular momentum must change as one moves from the outer, well ionized regions of the disk, into the region of the dead zone, where MRI turbulence will be damped and the bulk of the angular momentum flow is contingent on wind, and or Hall term transfer.  
 Disk winds do not physically act like turbulent viscosity - the disk does not spread radially outward under the action of a wind but is advocated inwards.
 Nevertheless, for modelling purposes, it is useful to consider an effective alpha parameter that characterizes the magnitude of angular 
 momentum transport out of the disk by an MHD wind.  
 If we designate an effective $\alpha_{eff}$ parameter for the disk, then one may write $\alpha_{eff} = \alpha_{SS} + \alpha_{wind} $, such that within the disk's dead zone $\alpha_{eff} \simeq \alpha_{wind}$.    
 The existence of a dead zone, expected from the basic physics of disk 
 ionization and the MRI instability, suggests that dust may more rapidly settle to the mid plane within $r_{DZ}$. 
Values of $\alpha_{SS}$ may drop to values lower than $10^{-4}$ in such regions.

\subsection{ Chemistry of evolving disks }

\subsubsection{Gas}

As the disk evolves, its changing physical structure is imprinted on its evolving chemistry. Of principle importance is the reduction of gas temperature and increasing ionization as the disk accretion rate decreases. Because reactions between ions and neutrals lack an activation barrier, the ionization rate plays an important role in dictating the rate of reaction for many gas phase reactions \citep{Eistrup2016}. As the disk surface density drops, and ionizing radiation can more easily penetrate to deeper regions of the disk, driving the chemical system to a (mathematically) steady state - where molecular abundances no longer change with time - more quickly. This steady state differs from the thermodynamic equilibrium solution for a set of reactions, whose final molecular abundances are dictated by Gibbs free energy minimization in that steady states are not necessarily global minima for the Gibbs free energy.  

Generally speaking, the gas changes its chemical structure through only a few chemical pathways. They are: freeze out onto and sublimation off of grain surfaces, neutral-ion gas phase reactions, neutral-neutral reactions on grain surfaces, and neutral-neutral gas phase reactions. The rates of each of these reaction pathways sensitively depend on the temperature, density, and ionizing flux of the disk's gas.

\begin{figure*}
\centering
\includegraphics[width=\textwidth]{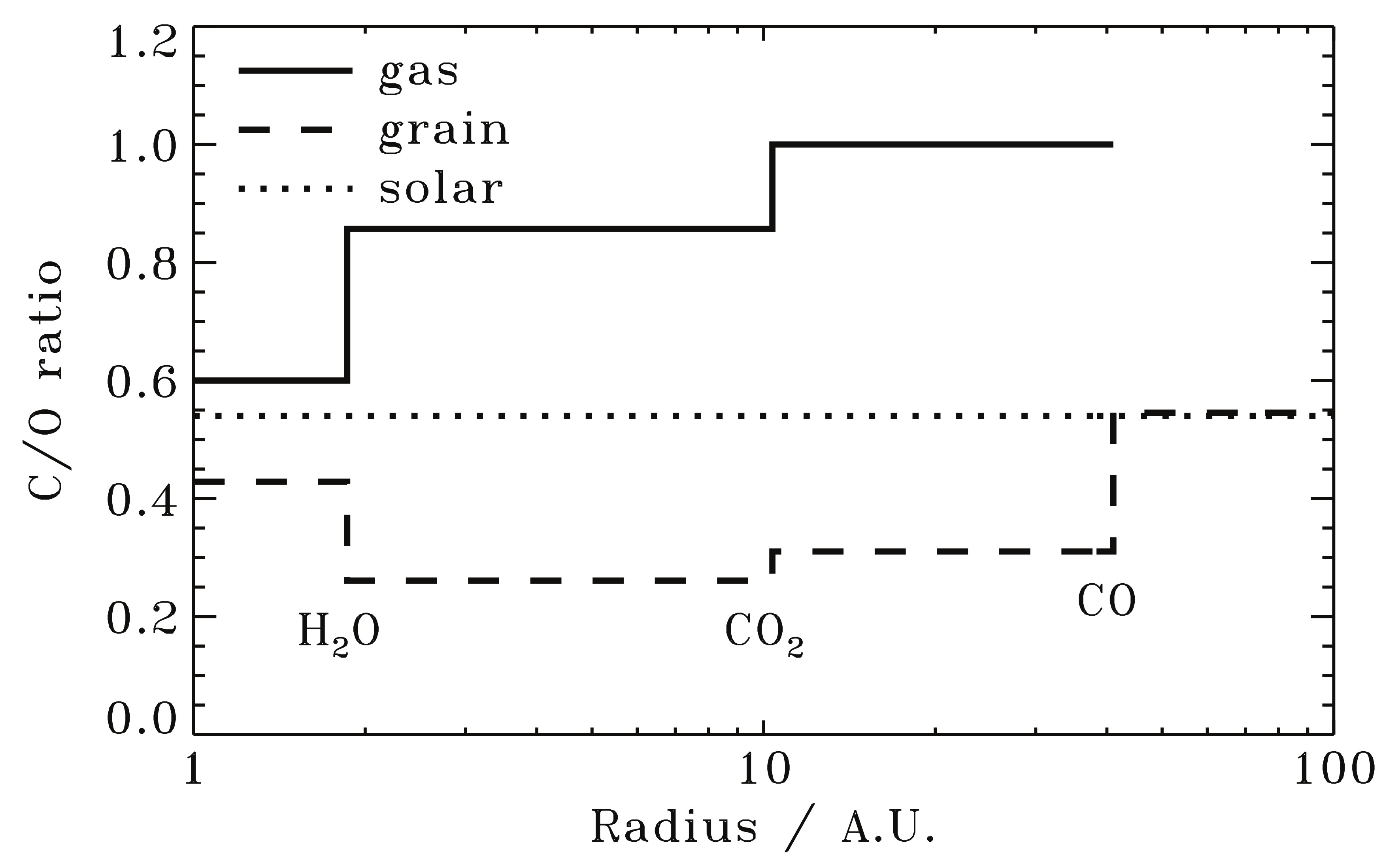}
\caption{The distribution of carbon and oxygen through a protoplanetary disk as shown by the carbon-to-oxygen ratio (C/O). The major jumps of C/O result from the freeze out of volatile H$_2$O, CO$_2$, and CO at their respective ice lines. Figure from \"{O}berg et al. (2011), ApJ, 743, L16. Reproduced with permission \textcopyright AAS.}
\label{fig:06}
\end{figure*}

Figure \ref{fig:06} (from \cite{Oberg11}) illustrates a well known, elegantly simple model of the elemental distribution through a disk. It shows the ratio of the total carbon and total oxygen elements (counting the most abundant molecules), known as the `carbon-to-oxygen ratio' (C/O), for gases and solids. At the ice lines of H$_2$O, CO$_2$ and CO, C/O changes as particular volatiles freezes onto dust grains. This process is dependent on the local gas temperature, so as the temperature of the gas cools the location of the ice lines (and their jumps in C/O) will move inward. In this model, the only chemical process that is taken into account is the freeze out of volatiles onto grains, which is balanced by their sublimation. In reality once a gas species has frozen onto a grain, it can be chemically processed while in the ice phase. This can be particularly important for the production of molecules like methanol which is produced through the hydrogenation of frozen CO \citep{Walsh2014}.

While these surface reactions will not generally change the C/O of the ices, gas phase reactions can have an impact on both the gas and solid C/O depending on where the reaction occurs. For example, CO has an exceptionally low freeze out temperature ($\sim 20$ K), and hence stays in the gas phase over a wide range of radii ($< 40$ AU, depending on the disk model). In its gas phase, it will most readily react with ions like He$^+$. This reaction results in its dissociation, leading to an ionized carbon atom, an oxygen atom, and a neutralized helium atom. The oxygen atom will quickly react with H$_2$ into water and freeze out onto the grains if the reaction occurs beyond the water ice line. The ionized carbon will also react with H$_2$ to produce CH$^+$, then through successive reaction with H$_2$ react into CH$_4$ which, if the reaction occurs outside of the methane ice line, will freeze out onto the grains \citep{Walsh2015}. Since the products of this route of CO destruction have different freeze out temperatures (CH$_4$ being lower than H$_2$O) the resulting C/O in both the gas and ice will be different depending on the radii where the reaction occurs. The rate of these reactions critically depend on the local ionization rate and temperature of the gas, hence their chemical evolution must be calculated in conjunction with the physical evolution of the disk.

It is particularly important to constrain the possible outcomes of chemical evolution because of its link to the way we interpret observations. For example CO is generally used as a tracer for the total gas in a protoplanetary disk, because it remains in the gas phase over a large radius, and its collisional excitation with H$_2$ is well understood from observational studies of molecular clouds. While it is assumed that CO should have the same global abundance in disks as it does in molecular clouds ($\sim 10^{-4}$ by number relative to hydrogen), observations of disks have suggested that it does not \citep{Dutrey1994,Bergin2013,McClure2016}. Multiple physical \citep{Salyk2008,Krijt2016,Xu2017} and chemical \citep{Bergin2014,Eistrup2016,Yu2016} methods of depleting the CO have been proposed, however this remains an open problem in protoplanetary disk science.

\subsubsection{Solids}

Left out of the above discussion is the chemical composition of the grains (often called the refractory material) on which grain chemistry occurs. This is because it is generally assumed that the grains condense very early in the disk lifetime. This assumption is supported by the fast ($\sim$ hour) condensation rate timescales that have been observed in lab experiments \citep{Toppani2006}.  Because of these fast reaction times (taking place faster than the disk's viscous evolution timescale), equilibrium chemistry methods can be used to compute the chemical abundance of the refractory material.

Equilibrium chemistry utilizes the thermodynamic result that a chemical system in equilibrium will have its total Gibbs free energy minimized. The Gibbs free energy of the system can be expressed as,

\begin{equation}
G_T = \sum_i^N X_i(G_i^0 + RT\log X_i),
\label{eq:Sol01}
\end{equation}

\noindent for a set of $N$ molecules, each with mole fraction $X_i$, Gibbs energy of formation $G_i^0$ at a temperature $T$ \citep{Pignatale2011,APC16a}. A second restriction is that the total number of elements:

\begin{equation}
\sum_i^N a_{ij} X_i = b_j \quad (j=1,2,...,m),
\label{eq:Sol02}
\end{equation}

\noindent where $m$ is the total number of elements in the system and $b_j$ are their initial mole fraction. $a_{ij}$ is the number of the $j^{th}$ element present in the $i^{th}$ molecule \citep{Pignatale2011,APC16a}.

Solving the equations \ref{eq:Sol01} and \ref{eq:Sol02} is generally done with a commercially available software package {\it HSC}, which
has been used in several astrophysical settings (eg. \cite{Pasek2005, Bond2010, Pignatale2011, Elser2012, Moriarty14, APC16a}).

\cite{APC16a} split the majority of refractories into two primary families: mantle, and core. Core materials are iron and nickel refractories that would settle to the core of a differentiated planet. While mantle materials are silicate, aluminium, and magnesium refractories that would end up in the mantle of a differentiated planet. They report that the primary core materials are iron (Fe), troilite (FeS), fayalite (Fe$_2$SiO$_4$), and ferrosilite (FeSiO$_3$). The primary mantle materials are enstatite (MgSiO$_3$), forsterite (Mg$_2$SiO$_4$), diopside (CaMgSi$_2$O$_6$), gehlenite (Ca$_2$Al$_2$SiO$_7$), and hibonite (CaAl$_{12}$O$_{19}$).

A third solid that can be tracked with equilibrium models are ices, however thermochemical data in equilibrium software is often incomplete for astrophysically relevant ices like CO$_2$ and CO. As a result, the only ice that is generally discussed in equilibrium contexts is water. 

\begin{figure*}
\centering
\includegraphics[width=\textwidth]{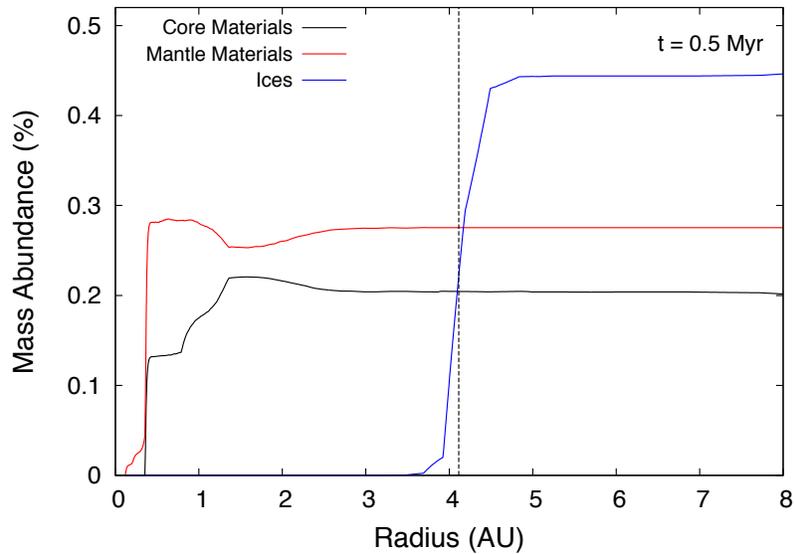}
\caption{A snapshot at t=0.5 Myr of the radial distribution of cumulative solid materials derived from a thermochemical equilibrium calculation in an evolving disk. Individual silicate and iron based refractories are combined into the more general `core' and `mantle' materials. Where the former will typically end up in the core of a differentiated planet, while the latter would end up in the mantle of a differentiated planet. The two phases of water, solid and gas, are found predominately outward and inward of the water ice line (dotted line) respectively. Figure reproduced from \citet{APC16a}, MNRAS, 464, 428.}
\label{fig:12x}
\end{figure*}

In Figure \ref{fig:12x} (from \cite{APC16a}) we show the radial dependence of the core, mantle, and ice material at the disk mid plane, early on in the disk lifetime. The mid-plane solid abundances are quantitatively similar
to those found by  \cite{Bond2010} and \cite{Elser2012}, who also performed equilibrium chemistry calculations on a disc of solar abundance.  Over most of the disk there is little variation between the abundance of core and mantle material, apart from the inner ($< 1$ AU) regions of the disk. In these high temperature regions of the disk, more complex core material like fayalite are less energetically favourable, hence the extra silicon is available to produce more mangle material like enstatite and forsterite \citep{APC16a}. As the disk evolves these features move inward with the inward motion of the gas and dust. The abundance of ices show the largest variation, due exclusively to the location of the water ice line (vertical dotted line in Figure \ref{fig:12x}). This feature likewise moves inward as the disk ages. The large variation in available ice abundance during the formation of planetesimals implies that the resulting composition (and hence structure, see below) will depend on where the initial planet core accretes.

\subsubsection{Dust radial drift}

As outlined in the chapter by Andrews and Birnstiel, the solid component of the disk (ie. the dust) evolves radially either due to turbulence (when the Stokes number is much smaller than one) or by radial drift (when the Stokes number is near one). This radial drift occurs when a grain is sufficiently large that its dynamics decouples from the gas. When this happens it is no longer affected by gas pressure, and begins to orbit faster than the gas (in Keplarian orbits). This velocity differences produces a head wind on the grain which drains angular momentum, moving the grain to a smaller orbit. The speed at which it drifts is related to the Stokes number (St) through \citep{Weidenschilling1977}:
\begin{equation}
u_{drift} = -\frac{2 u_\eta}{St + St^{-1}},
\end{equation}
where $u_\eta$ is the difference between the dust orbital speed and the gas, and hence the drift is most efficient for grains with St $=1$.

This drift has important implications for the chemical structure of the disk because dust plays an important role in setting both the opacity of higher radiation (hence impacting the ionization state of the gas), and the rate of gas freeze out (by providing the sites necessary to freeze). Dust radial drift can also act as a transport mechanism for ices because it can transport frozen species across their respective ice lines, enhancing the abundance of gases at these locations.

It has long been known that ice lines could play an important role as the sites for rapid particle growth by  condensation.  The original suggestion by \cite{StevensonLunine1988}
was that there could be a considerable enhancement of material at an ice line from vapour that diffuses from the inner part of the disk.  The enhancement of vapour by the evaporation of materials moving inwards across the ice lines was addressed by \cite{CuzziZahnle2004}.  
Recently, the transport properties of dust grains have been investigated in the context of volatile transport across ice lines \citep{Stammler2017,Booth2017,Bosman2017b}. These works have demonstrated that radial drift is sufficiently fast to transport ices across their ice line before sublimation returns the volatile back to the gaseous state. \cite{Bosman2017b} report that grains with a Stokes number of unity (the most susceptible to radial drift) have drift timescales of approximately 100 yr. The rate of sublimation (per volume) is given by \citep{Bosman2017b}: 
\begin{equation}
f_{sub} = p_x\sigma_{dust}n_{grain}N_{act}\exp\left[-\frac{E_{bind}}{kT}\right],
\label{sq:sub01}
\end{equation}
where $p_x$ is a prefactor, $\sigma_{dust}$ is the surface are of the dust, $n_{grain}$ is the number density of grains, $N_{act}$ is the number of ice layers available for sublimation (usually 2), $E_{bind}$ is the volatile's binding energy, and $T$ is the temperature of the dust and gas (assumed to be the same). Using the values from \cite{Bosman2017b}, and assuming a Stokes number of unity we estimate a reaction time ($n_{H2O,ice}/f_{sub}$) of a few 100 yr for grains crossing the water ice line. This implies that as the grain crosses the ice line, it does not immediately lose its ice layer. However, because the sublimation rate scales as $\exp(-1/T)$, the grains do not travel far inward of the ice line before losing all of its ice. In fact an increase of only 30 K (radial change of about 0.1 AU for the disk model in \cite{Bosman2017b}) results in a reduction in the sublimation time of 2 orders of magnitude! Once in the gas these enhancements spread out through diffusion, delivering some of the volatile outward of its ice line where it refreezes onto grains - continuing a cycle of freeze out, transport, sublimation, and diffusion (see also \cite{RosJohansen2013}).  The end result is an enhancement of the C/O within ice lines in the planet forming regions of the disk \citep{Booth2017}, possibly being imprinted into the atmosphere of a forming planet.

Radial drift also plays a role in dictating the ionization state of the gas, because the dust (which contributes highly to the disk opacity) is rapidly cleared from the outer regions of the disk. \cite{Crid16b} showed that  when this happens,  the region of the disk with low ionization rapidly shrinks, moving the outer edge of the dead zone inward (to $\sim 1$ AU) quickly (within 1 Myr). Because of its tendency to rapidly drive chemistry, a rapidly changing ionization structure will have an important impact on the chemical structure of the disk. This rapidly shrinking dead zone is at odds with the picture of a laminar accreting disk driven by disk winds (see above).  We note however,
that the ALMA observations clearly show that radial dust flow in disks is much slower than the current models 
predict  (see the chapter by Andrews and Birnstiel).

A fully self consistent treatment of the dust physics and photochemistry has not yet been undertaken because of technical challenge which include incorporating the radial movement of material in a chemical evolving system of equations. These complexities have begun to be incorporated by \cite{Bosman2017b} for a limited chemical network dedicated to the formation and destruction of CO$_2$ in disks. However without a full chemical network as seen in \cite{Walsh2014}, \cite{Helling2014}, \cite{Eistrup2016}, or \cite{Crid17} the detailed evolution of the C/O remains elusive.

\section{ Planet migration in inhomogenous disks - planet traps}

The essence of the standard theory of planet migration in smoothly varying disks was uncovered decades ago. One of the key ideas is that planets can induce the launch of waves at Lindbland resonances in the disk where the forcing frequency of the planet equals the epicyclic frequency of fluids motions oscillating around their guiding Kepler orbits (eg. \cite{GoldreichTremaine1979,Ward1986,LinPapaloizou1986}.  For torques exerted by the outer and inner Lindblad resonances the corresponding resonances are at epicyclic frequencies of $ \kappa (r) = [ (m / m + 1) ; (m / m-1) ] \Omega_p $  respectively, for waves with azimuthal wave numbers $m$.  These exert torques in opposite directions, with the outer resonance (outward angular momentum transport) being the larger as it is closer to the planet due to the gas' sub-Keplerian orbital velocity - a consequence of the local gas pressure gradient. Thus, the net torque generally leads to the planet losing angular momentum. 
  
In these models, if the planet mass is small enough, the disc response is linear. The migration rate is then proportional to the planet and disc masses, independent of the viscosity and weakly dependent on the disc surface density and temperature profiles. This is the so-called Type I migration \citep{Ward1997}.  If unopposed, these torques would push an Earth mass orbiting at 1AU into the central star within $\sim 10^5$ years, a timescale that gets shorter as the mass of the planet increases.  More detailed analysis  of the magnitude of the Lindbland torque that fits the results of 2D numerical simulations to disk models that include the effects of smooth power-law behaviour of the disk surface density $\Sigma \propto r^{- s} $ and the disk temperature $ T \propto r^{- \beta}$ were \citep{Paardekooper2010};

\begin{equation}
\Gamma_L / \gamma \Gamma_o = -2.5 - 1.7 \beta + 0.1 s 
\end{equation}  

\noindent where $\gamma$ is the adiabatic index of the gas and the torque $\Gamma_o = (q / h)^2 \Sigma_p r_p^4 \Omega_p$ can be readily  derived by calculating the change in the angular momentum of a fluid element perturbed in passing a planet
of mass $M_p$ for quantities evaluated at the position of the planet.   

As one comes in closer to the planet and the co-rotation region,  parcels of gas undergo nonlinear perturbations and move along "horseshoe" orbits wherein cooler, higher  angular momentum fluid to the planet's exterior undergoes a sharp U- turn in front of the planet and swapped into an inner orbit where gas is hotter and has less angular momentum.  
The opposite occurs for inner fluid moving outwards in the U.  Since entropy must be conserved during these motions,  a density enhancement near the planet develops resulting in an  outward net torque (\cite{Ward1997},  Nelson's chapter).  Similar effects arise from the conservation of vortensity during these motions.  The resulting co-rotation torque was also computed by \cite{Paardekooper2010} and takes the form

\begin{equation}
\Gamma_{HS} / \gamma \Gamma_o = 1.1 ({3 \over 2} - s) + 7.9 (\zeta / \gamma) 
\end{equation}

\noindent where $\zeta = \beta - (\gamma - 1) s$ is the power law exponent for the entropy profile of the gas, and 
the first and second terms address the entropy and vortensity effects.  The total torque that a  low mass planet undergoes dying Type I migration is  the sum of these two the Lindblad and horseshoe co-rotation torques; $\Gamma = \Gamma_L + \Gamma_{HS}$.    

The values of the power law indices appearing in this combined formula  have been worked out for \cite{Chambers2009} disk
models (see \cite{Crid16a}) for an adiabatic index of $\gamma = 1.4$. As has been observed by several authors the net torque is outward in the viscous regime and inward in the radiative regime.  This implies that there is a radius of net zero torque $\Gamma = 0$  at the heat transition radius $r_{HT}$ discussed above  \citep{Lyra2010,HP11,Dittkrist2014}.   Thus a low mass planet can be trapped at the heat transition, and this cuts off its rapid inward migration.   A planetary core trapped there moves with the trap, and accretes materials that are at the trap position as the latter moves through the disk.   The evolution of $r_{HT}$ with time is governed by the disk evolution equation discussed above, and in particular depends on the accretion rate, which falls as a function of time as the column density falls, and later as the disk undergoes photo evaporation.   These motions are much slower functions of time than the Type I migration time scale because they reflect the slower viscous evolution of the disk which is responsible for the falling accretion rate $\dot M_a $.   There is a range of masses that can be trapped in regions of zero net torque.  Detailed simulations by \cite{ColemaneNelson2016} examined the growth of planets that grow in traps caused by radial variations in the 
disk.  These results showed that null points for the torque can trap planets up to 10s of Earth masses.  For the more massive planetary cores, being released from the trap will, with rather little additional mass accretion, result in gap opening and the transition to slow  Type II migration (see below).  

Two other general types of inhomogeneities in disks are possible.   Noting that the dead zone must have very low levels of MRI turbulence, while the active regions at larger radii can support active MRI, there is a discontinuity in the dust scale height as one proceeds through the outer dead zone radius .  For disk radii   $ r \le r_{DZ} $,  dust will rapidly settle into the disk mid plane whereas at $ r \ge r_{DZ}$, turbulence will keep the dust stirred up to higher scale heights.   This means that radiation from the star will see a "wall of dust" at $r_{DZ} $ which will reflect and, as for a garden bed in front of a sunlit wall, will be heated by the back-scattered radiation.   This alters the 
temperature profile of the gas at smaller radii, in such a way as to create a planet trap \citep{HP11,Crid16a}.

The position of the dead zone, as we have seen, depends on the ionization of the disk and hence is directly connected to ionization driven disk chemistry. A planet trapped at the dead zone radius migrate with the trap's evolution, acquiring a composition reflecting the disk materials encountered as the trap moves inwards through the disk. 

The third generic type of disk inhomogeneity  is  the entire class of ice lines that result from the freezing out of various chemical species on grains.  As has already been noted, three of the potentially most important ice lines are those of water, CO, and CO$_2$.  In its most general form,  astrochemical models for  the disk predict the distribution of ices (eg. water).  These results can then be used to compute the change in dust opacity in the disk.  This opacity change has a direct effect on the temperature profile across the ice line, which in turn sets the direction of the torque by the temperature dependence of equation 9. In order to have a sufficient large change in the opacity across an ice line, one can anticipate that the relevant volatile must be abundant - as is the case for water \citep{Miyake1993}.   Water ice lines have accordingly been suggested as trapping points for planets \citep{IdaLin2008b,HP11}.

Why does an ice line act as a potential trap?  At the ice line, the opacity $\kappa$ is reduced as the dust
grains are coated with ice, and the associated cooling rate of the gas is increased (since the cooling rate $\propto 1/ \kappa $; \citep{Bitsch2013}). Coupled to the cooling rates, the local temperature and thus the disk scale height, is reduced ($H = c_s \Omega  \propto T^{1/2} $) . Since the disk accretion rate is  constant  across the disk at any instant, and because the viscosity is dependent on the local gas temperature ($\nu \propto c_sH \propto T  $), a reduced temperature results in an enhancement in the local surface density at the ice line to maintain a constant mass accretion. Both the modified temperature and density gradients impact the net torque, resulting in a trap.

Detailed analysis of opacity effects at ice lines indicate that water is sufficiently abundant $ 1.5 \times  10^{-4} $ 
molecules per H) to trap planets at its ice line due to an opacity transition. Volatiles that have mass abundances lower than a factor of $\sim$40 with respect to water do not result in a sufficiently strong opacity transition to trap planets in a disk that is viscously heated as shown in numerical simulations (Cridland, Pudritz, \& Alessi, in press).   These results suggest that CO, while sufficiently abundant, also does not trap planets at its ice line. This is because its ice line is in the outer parts of the disk where heating is dominated by irradiation by the host star rather than viscosity. In this heating regime, the midplane temperature is not as dependent on the dust opacity, and hence an opacity transition does not cause a strong change in the temperature profile as seen at the water ice line. Like water, the CO$_2$ ice line occurs in the viscously heated part of the disk and hence could act as a trap. However it is not easily produced in the gas phase, and hence its availability as a trap depends on your choice of initial conditions (dark cloud vs. diffuse ISM chemistry) and chemical network (gas only vs. gas-grain chemistry).

We reach an interesting and important conclusion about planetary migration that has direct consequences for both planet
formation and composition.   Low mass planetary embryos move along with traps that move inwards through the disk
as it evolves on a viscous time scale.  Planets eventually break away from their traps when they become sufficiently massive as shown in numerical simulations of \citep{ColemanNelson2014} - typically up to 10 Earth masses.    
The solid materials accreted during this time reflect the composition of the evolving disk visited by the relevant trap.
The heat transition, being typically the furthest out in the disk, is beyond the ice line and so planetary cores can be
expected to have a strong contributions of ices \citep{APC16a}.  A core building at the ice line would be expected
to have less ice, and a dead zone, which is often inside the ice lines, would be expected to have a very small ice content.
Detailed simulations bear out these general results (see subsection on solids).  

The dynamics of embryos as they approach planet traps has not yet been investigated in any detail.   One
anticipates that a rapidly building core may scatter incoming materials.   As a planet builds in a trap, one 
also anticipates that a chain of embryos will come into mean motion resonances with the traps, resulting 
in a series of planets undergoing accretion associated with each trap.  

As the mass of planets increase, the angular momentum exchange between disk and planet can lead to the 
opening of a gap, a process called Type II migration \citep{Ward1997}.  In this regime,  a balance is established between 
the tidal torque which tends to open a gap (inner material has angular momentum removed and moves inward, which is then
transferred via the planet to the outer martial which moves outwards) - and viscous torques which always act to fill in 
a gap.   

 The `gap opening' mass requires the planet's Hill radius to be larger than the pressure scale height at the planet's location, otherwise the gap will be closed by gas pressure. A second requirement is that the torques from the planet on the disk exceeds the torques caused by viscous stress. These requirements are summarized by, \citep{LinPapaloizou1993,HP11},
 
 \begin{equation}
{M_p \over M_* } =  min[ 3 h_p^3, \sqrt{ 40 \alpha_{turb} h_p^5} ]
\end{equation} 

\noindent where $h_p = H / r_p$. The depth and width of the gap depend on this balance \citep{PapaloizouLin1984}.  The 
Lindblad resonances which drive the disk-planet angular momentum exchange fall into opening gap and therefore the migration rates are drastically reduced compared to Type I.  The planet then becomes locked to disk migration at the radial velocity of $u_r = \nu / r $, and the disk is essentially split into an inner and outer region.   This result is based on the assumption that gas doesn't enter the gap once formed.  This, is  in fact to simplistic a view since horseshoe orbits can readily facilitate a flow
through the gap.   Numerical studies carried
out by \cite{CridaMorbidelli2007} and \cite{Edgar2008}, have recently been generalized by \cite{Duffell2014} who showed that Type II migration can, on this
basis, be faster or slower than the viscous rate depending on disk parameters such as the turbulent Mach number.   

One of the main results of the early population synthesis studies  arose when the effects of Type I planetary 
migration  \citep{IdaLin2008} were considered.  Synthesis studies of planets migrating in evolving,  standard, Shakura-Sunyaev smooth disks showed that rapid loss of such bodies occurs within $10^5$ yrs.   The model introduced a parameter - a slowing down factor - needed in order  to match predicted and observed populations in the M-a diagram.  The result was that standard migration in smooth disks needed to by slowed down by a factor of 30-300 (see Nelson's review) - i.e. - slowed to speeds more reflecting viscous evolution of the disk.  The theory of planet traps, sketched above, provides a physical solution for this problem.  

The results of planet migration on planetary populations and their evolutionary tracks in the M-a diagram are discussed in 
the following section.   We emphasize that these are a consequence of a treatment of co-rotation torques that depend on
disk viscosity. In MHD wind driven regions (eg. the dead zone)  disks may be considered to be inviscid and there, co-rotation torques will arise from MHD disk winds (\cite{McNally2017}, review in Nelson's chapter).  Here, the shape of the horseshoe orbit region near the planet can be modified by the winds, leading to a more "history-dependent" evolution of the horseshoe torque.  

\section{ Planet formation \& composition }

The process of planet formation is complex and involves growth from interstellar dust ($\sim \mu$m) up to Jupiter-sized planets ($\sim 10^5$ km), a range of 14 order of magnitude! Of course modelling the complete process is be impossible, and hence most models begin or end at an intermediate stage called a planetesimal (size between 1-100 km). 

Growth from interstellar dust to planetesimals requires crossing the `meter-barrier' - a size where fragmentation and radial drift destroys the object faster than it can grow (see chapter by Andrews and Birnstiel) . A method for circumventing the meter-barrier is through the streaming instability \citep{YoudinShu2002,YoudinGoodman2005,Raettig2015,Schafer2017}. Simulations
by  \cite{Simon2016} found that the resulting planetesimal mass distribution scales as $ dN/dM_p \propto M_p^{-1/6 \pm 0.1}$ up to objects
on the scale of the large asteroids or Kuiper Belt objects.  Hydrodynamic simulations of the streaming instability can rapidly build up a population of planetesimals available for further growth \citep{Johansen2007,Schafer2017}.

There are two primary methods of solid accretion to grow planets from the initial planetary embryo: planetesimal accretion and pebble accretion. The primary difference between planetesimal and pebble accretion is the size of the accreting mass. Pebble accretion assumes that the planetary embryo is accreting $\sim$cm sized `pebbles' rather than km-sized planetesimals. As as result the effective gravitational focusing  is much higher for pebbles because their speed relative to the embryo is much smaller than planetesimals \citep{Bitsch2015}.

\begin{figure*}
\subfloat[Individual planet formation tracks representative of 3 different planet traps.]{
\includegraphics[width=0.5\textwidth]{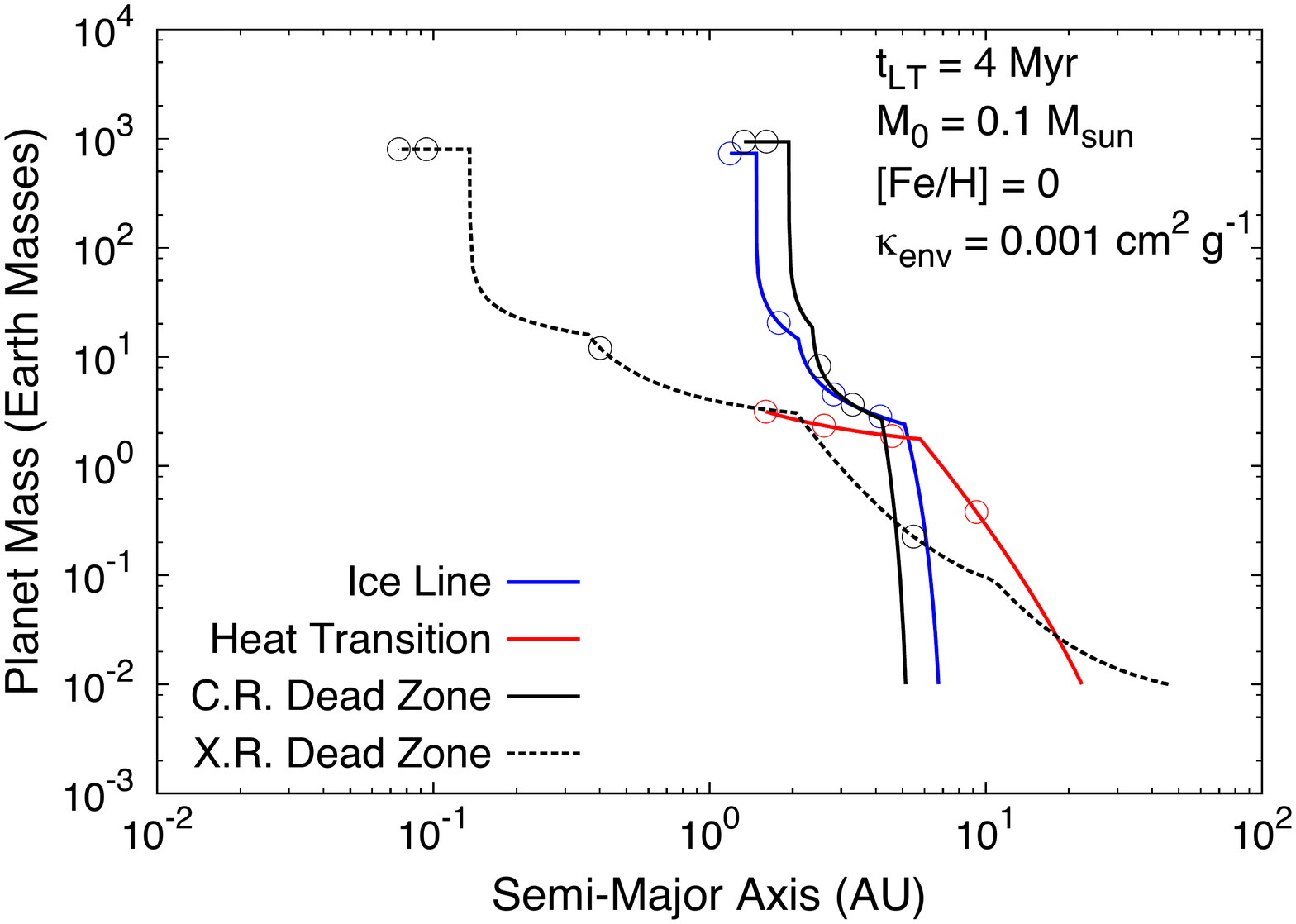}
\label{fig:13a}
}
\subfloat[Full synthesized population of planets.  The colour code is for planets formed on the three different planet traps ]{
\includegraphics[width=0.5\textwidth]{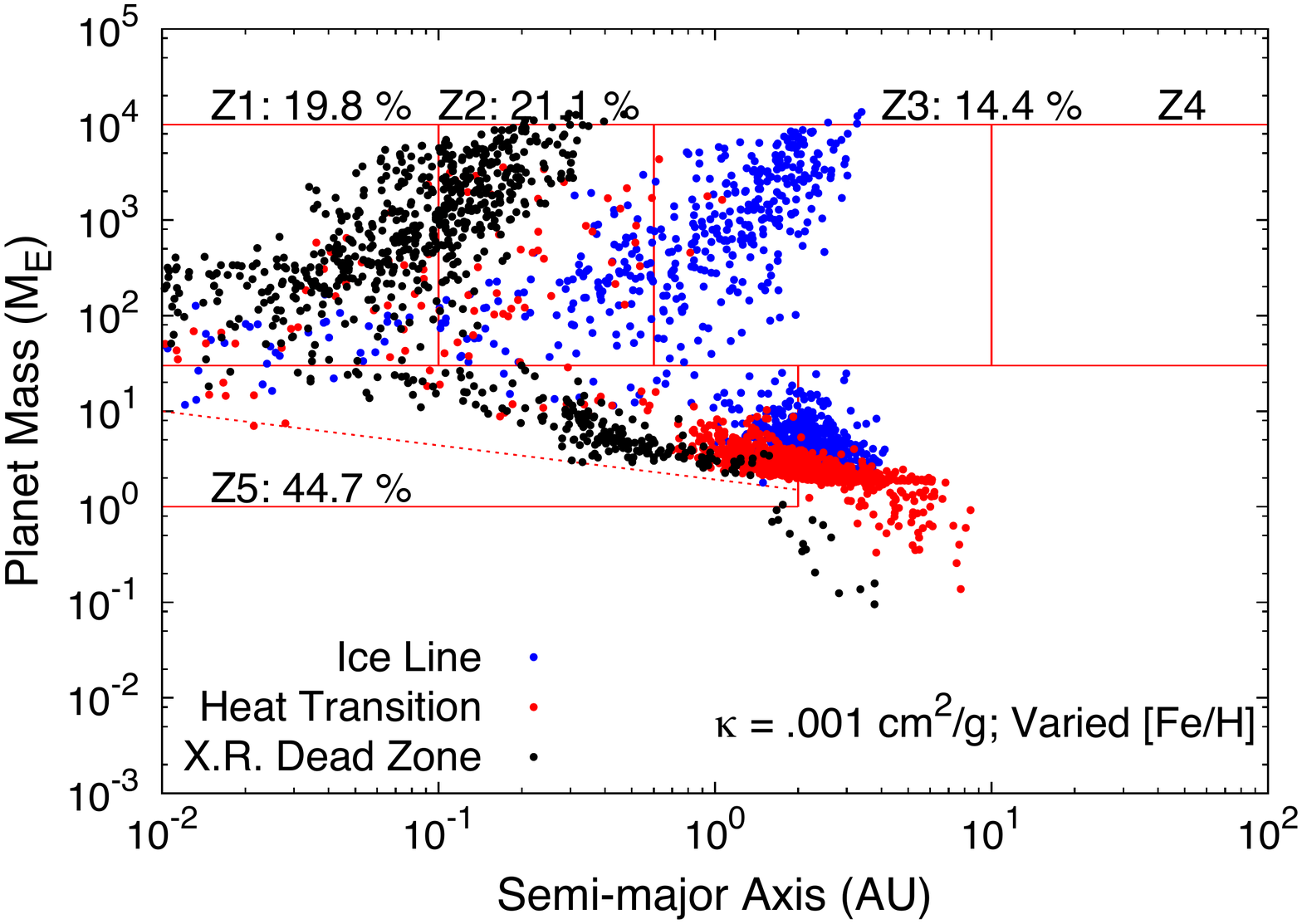}
\label{fig:13b}
}
\caption{Sampling many planet formation tracks (see a) with different disk initial conditions and physical parameters is used to construct a synthetic population of planets (see b). Comparing the occurrence rate of different types of planets to the observed population helps to constrain the physical properties of the prototplanetary disk and the growing planet which lead to the observed population of planets. Figures reproduced from \citet{Alessi2018}, arXiv:1804.01148 (submitted to MNRAS).}
\label{fig:13}
\end{figure*}

Both of these mechanisms grow a planetary core with a mass of between 5-10 M$_\oplus$, at which point the planetary core begins to accrete its gaseous atmosphere (see the chapter by Mordasini for details). Gas accretion occurs in two phases - a slow phase where the planetary envelope remains connected to the surrounding protoplanetary disk gas, and a fast phase where the planetary envelope decouples from the disk. These two phases, along with the initial phase of solid accretion can be seen in Figure \ref{fig:13a}.

In Figure \ref{fig:13a} we show individual formation tracks (the evolution of a planet through the mass-semi-major axis diagram) for a typical growing planet in three
traps indicated by the colour code.  The size and evolution of dead zones in disks depends on the type of 
disk ionization that dominates.  In this model, a constant dust to gas ratio is assumed for the gas.
 Shown in the figure are tracks for planets in dead zone traps,  for disks ionized  either by cosmic rays (C-R deadzone) or X-rays (X-R deadzone).  
Planets generally evolve from right to left, and bottom to top of the diagram following the traps as they move slowly inward through the disk
on $10^6$ year type of time scales.   Planet evolution tracks start in planet traps as the embryos are assumed to migrate very quickly until
they encounter a trap.   They then  begin their growth in the Oligarchic stage (see \cite{KokiboIda2002} and \cite{IdaLin2004}) by accreting solids until the majority of solids have been cleared out of the planet's feeding zone. The timescale associated with this accretion ($\sim 10^5$ yr) is much shorter than the migration timescale ($\sim 10^6$ yr) if the planet begins close ($< 10$ AU) to the host star. So the planet evolves nearly vertically on the diagram (see the blue and black curves). Farther out in the disk, the solid accretion rate is comparable to the migration rate, and the planets evolve more diagonally (red and black dashed line). Once the rate of solid accretion drops, the core can cool to begin accreting gas, and enters into a phase of slow gas accretion. 

Beginning in this slow phase of accretion, the gas is first accreted into an envelope that remains connected with the surrounding disk, but slowly contracts. This contraction is limited by the rate that the envelope can radiate its energy away, and hence contracts on the Kelvin-Helmholtz timescale (\cite{IdaLin2004}, or see the chapter by Mordasini). In this phase the contraction occurs at a slower timescale than the (trapped) migration timescale ($\sim 10^6$ yr), and the planet evolves nearly horizontally across the diagram. 

For masses exceeding a critical mass $M_p > M_{\rm{c,crit}}$, gravitational instability ensues and the planet's gas envelope grows by accretion from 
the disk on the Kelvin-Helmholtz timescale \citep{Ikoma2000},
\begin{equation} 
\tau_{KH} \simeq 10^c \, \textrm{yr}\left(\frac{M_p}{M_\oplus}\right)^{-d}\;.\label{Gas_Accretion}
\end{equation}
The values of parameters $c$ and $d$ in the Kelvin-Helmholtz timescale are physically linked to the opacity of the 
accreting planet's atmosphere, $\kappa_{\rm{env}}$.   This is included in the model by using the fits shown in \citet{Mordasini2014}, that relate results of a numerical model of gas accretion to the Kelvin-Helmholtz parameters for a range of envelope opacities of $10^{-3}-10^{-1}$ cm$^2$ g$^{-1}$. The fit given for the Kelvin-Helmholtz $c$ parameter is,
$ c = 10.7 + \log_{10}\left(\frac{\kappa_{\rm{env}}}{1\,\rm{cm}^2\,\rm{g}^{-1}}\right)\;.\label{KHc}$
The Kelvin-Helmholtz $d$ parameter has a more complicated dependence on envelope opacity, ranging from $\approx$1.8-2.4 over the range of $\kappa_{\rm{env}}$ considered,
the details being given by a  piecewise-linear function shown in \citet{Mordasini2014}.  

Once the planet becomes sufficiently massive the contraction becomes rapid and the envelope decouples from the surrounding disk. In this rapid phase the gas accretion timescale 
($\leq 10^4$ yr) is much lower than the migration timescale, and hence the planet again evolves nearly vertically. 
Clearly this unstable phase must not last long, and there have been multiple suggests to limit the rate of gas accretion. One way is to simply place an upper limit on the mass of the planet. This limit is often set as a large ($\sim 50-100$) multiple of the gas opening mass (see \cite{HP13} and  \cite{APC16a}). Alternatively, the gas accretion can be limited by other accretion rates once a gap is opened in the disk. In these models, the gas accretion is limited by the Bondi accretion rate, or the global disk mass accretion rate (see above). Because the majority of the gas mass is accreted during this fast accretion phase, changing the gas accretion prescription could result in changes in the chemical composition of the gas as it accretes onto the planet. Whether these would be measurable changes has not yet been investigated.

In Figure \ref{fig:13b} we show results from a population synthesis of our formation model \citep{Alessi2018}. Each planet has evolved through different regions of the disk at different times, because they are the result of different disk initial conditions and parameters, and hence have accreted gas with potentially different chemical histories.   A key result from this population synthesis study is that a reasonable fit to the observed M-a diagram with its separated
hot and warm Jupiter populations, such as shown in this figure, required a low envelope opacity ( $\kappa_{env} \simeq 0.001$ cm$^2$ g$^{-1}$  )
that is three orders of magnitude smaller than the opacity of materials in the disk.
Higher envelope opacities lead to inefficient cooling and lower accretion rates,  
which means that the planets are dragged into the inner regions of the disk before they start accreting much 
mass.   This, in effect, washes out the warm Jupiter population.  

Another result seen in this population study is that the warm Jupiters are primarily formed in the ice line trap - a conclusion also reached by \cite{HP13} and
first suggested in \cite{IdaLin2008}. Of particular interest is that low mass super Earths do not appear in the innermost regions (less than 0.1 AU) of the M-a diagram.
This may be a consequence of the model in that dust does not undergo radial drift.  If it did, it may be possible to populate this region, but this remains to be computed.
\cite{Hasegawa2016} has indeed noted that another formation mechanism - such as terrestrial planet formation - may be needed to explain the  low mass  planets that end up in such close in orbits.   


With the kind of machinery that has been discussed here, one is in a position to compute the chemical composition of the material that is accreted during the planet formation process. These models are called either `end-to-end' or `chain' models because they link physical and chemical models together in succession. A generally construction of these end-to-end models is {\it disk model} $\rightarrow$ {\it chemical model} $\rightarrow$ {\it planet formation \& migration}. Variations and extensions of this general chain have been made, including a dust evolution model \citep{Crid16b} and a planetary atmosphere model which includes the generation of synthetic spectra \citep{Mordasini16}. The chemical model can be construct empirically \citep{Mordasini16}, through chemical equilibrium models \citep{APC16a}, or from time-dependent chemical models \citep{Crid17}, depending on the focus of the chain.

\subsection{Planet Cores} 

The physics of planetary interiors and cores depends on the equations of state (EOS) of the primary
materials such H, He, and materials made of water, silicates, and iron - as well as the overall chemical composition of the planet (see review \cite{Baraffe2014}) .  Knowledge of the first aspect of this problem relies on experimental studies of the properties of materials under high pressure, as well as by satellite probes of the density structures of the giant
planets such as Jupiter, recently achieved by the Juno mission \citep{Bolton2017}.
It is in the second aspect - the range in overall compositions of planets that give rise to the enormous diversity
to the M-R relations for planets.    The elemental composition of planets is directly connected to the materials
planets accreted as they were formed.   Thus, the structural properties of planets as seen in their M-R relation,
are a direct consequence of planet formation. 

\begin{figure*}
\centering
\includegraphics[width=\textwidth]{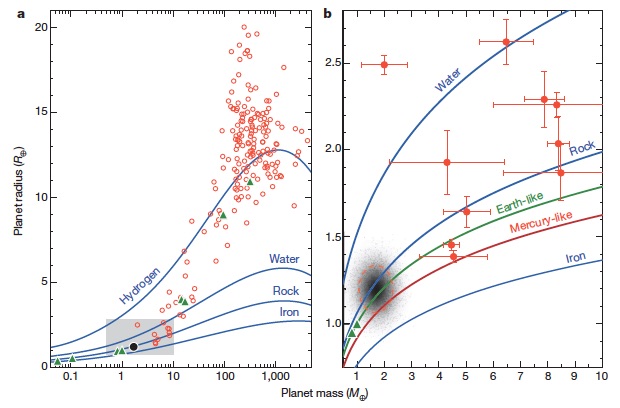}
\caption{The planet mass-radius (M-R) diagram for a set of Hot Jupiters and sub-Earths. In the left panel, the M-R curves denote planets that are made from pure hydrogen, water, rock and iron. The right panel is a blow up of the greyed region in the left panel, and additionally shows the M-R curves for Earth-like, and Mercury-like planets. Clearly there is a diverse set of compositions in the super-Earth's (right panel) as no internal model uniquely describes every planet. Likewise, Hot Jupiters (left panel, top right) can be highly irradiated, and hence are `puffier' than the hypothetical `pure hydrogen' planet. Figure from \citet{Howard2013}, Nature, 503, 381. Reproduced with permission \textcopyright Springer Nature.}
\label{fig:12}
\end{figure*}

The observations show that exoplanets are incredibly diverse.  In Figure \ref{fig:12}, left panel \citep{Howard2013}, we observe that planets with a given mass can have an enormous range of sizes.  Thus for the Jovian planets, (a few hundred Earth masses), one observes of a factor of 2 range in planetary radius for a given mass.    One also notes that there are 
 "inflated"  Jovian planets whose radii greatly exceed those predicted for a composition of pure hydrogen. 
 A variety of models have been proposed to explain such planets including stellar heating  \citep{ShowmanGuillot2002,Weiss2013} and heavy element gradients in the planetary interior that would decrease the rate of heat transport thereby slowing down the cooling and contraction of the planet \citep{ChabrierBaraffe2007}. Similarly, for Super Earth masses, there is a wide range of radii for any given mass again indicative of planets that are composed of a very wide range of materials. The right panel of Figure \ref{fig:12} is a blow  up of the Super Earth mass regime (1 - 10 Earth masses), and clearly shows the existence of planets with densities corresponding to rock-iron mixtures. 

The effects of heavy metal enrichment on planet structure were first carried out by \citep{ZapolskySalpeter1969}, for planets made of individual elements at $ T = 0 $.   If convective energy transport dominates in the planetary interior, the temperature gradient will be nearly adiabatic.  As is seen from Figure \ref{fig:12}a, increasing the heavy element content in giant planets leads to a decrease in their radii. Composition gradients within their interiors suppress their interior cooling, leading to models that are hotter than the adiabatic models \citep{LeconteChabrier2013}.  

The core accretion picture of Jovian planet formation predicts the existence of rocky, 10 Earth mass cores  (\cite{Mizuno1978,BodenheimerPollack1986,Pollack1996} - and chapter by D'Angelo and Lissauer)   What evidence is
there for this picture?    And what chemical state is the interior of the planet in?  If Jovian planets form by gravitational instability, then there would be no initial core.  If a core did form, it could erode away with time as it slowly dissolves in liquid metallic hydrogen \citep{Stevenson1985,Gonzalez2014} enriching the envelope above it.  

These fundamental questions were addressed  by one of the most significant experiments in planetary science over the last decade - the Juno spacecraft.  The mission goal  is to improve our understanding of the origin and evolution of Jupiter, the history of the solar system, and planetary system formation in general.   On August 27, 2016, this probe flew less than 5000 km over the equatorial cloud tops of Jupiter acquiring a wealth of measurements of the state of the planet's atmosphere, magnetic field, and interior structure \citep{Bolton2017}.  The gravity measurements, found by determining small deviations of the spacecraft trajectory due to low order harmonics ( $J_4, J_6$ in particular ) of Jupiter's gravitational field, were an order of magnitude more sensitive than any before it.   

The modelling of the planetary interior and comparison with the data \citep{Wahl2017} considered interior density profiles that are in hydrostatic equilibrium;  

\begin{equation}
\bigtriangledown P = \rho \bigtriangledown  \Phi
\end{equation}

\noindent where a barotropic pressure $P(\rho) $ corresponding to isentropic profiles constructed from various EOS is used.
Numerical simulations  of H-He mixtures from (\cite{MilitzerHubbard2013}, MH13) were employed.  Density functional theory molecular dynamics simulations are the best technique  for determining densities of hydrogen-helium mixtures over most of conditions in a giant planet ( $ P > 5 $ GPA).   

The results show that Jupiter has a core in the range of 6-25 Earth masses.   The results are generally consistent with the core accretion - collapse model \citep{Pollack1996}. The larger masses correspond to having a more dilute density profile in the core, equivalent to extending about 10 Earth masses of material out to 0.3- 0.5 Jovian radii,  $R_J $.,
This agrees with models that account for the dissolution of planetesimals \citep{Lozovsky2017} as the reason for 
a dilute core structure.   It is not known whether there is enough convective energy available to lift so much material.   The overall results clearly depend on exactly how the planet formed, and how mixing occurred during these early stages \citep{LeconteChabrier2012}.   The mass of the heavy elements in the envelope depends strongly on the EOS, with MH13 predicting 5-6 times solar heavy metal fraction in Jupiter.  

\begin{figure*}
\centering
\subfloat[A breakdown of the refractories that are accreted into a Super Earth with a final mass of 5.4 Earth masses, formed
at the heat transition trap $r_{HT}$, in a calculation with a disk life time of 3 Myr. Figure reproduced from \citet{APC16a}, MNRAS, 464, 428.]{
\includegraphics[width=0.5\textwidth]{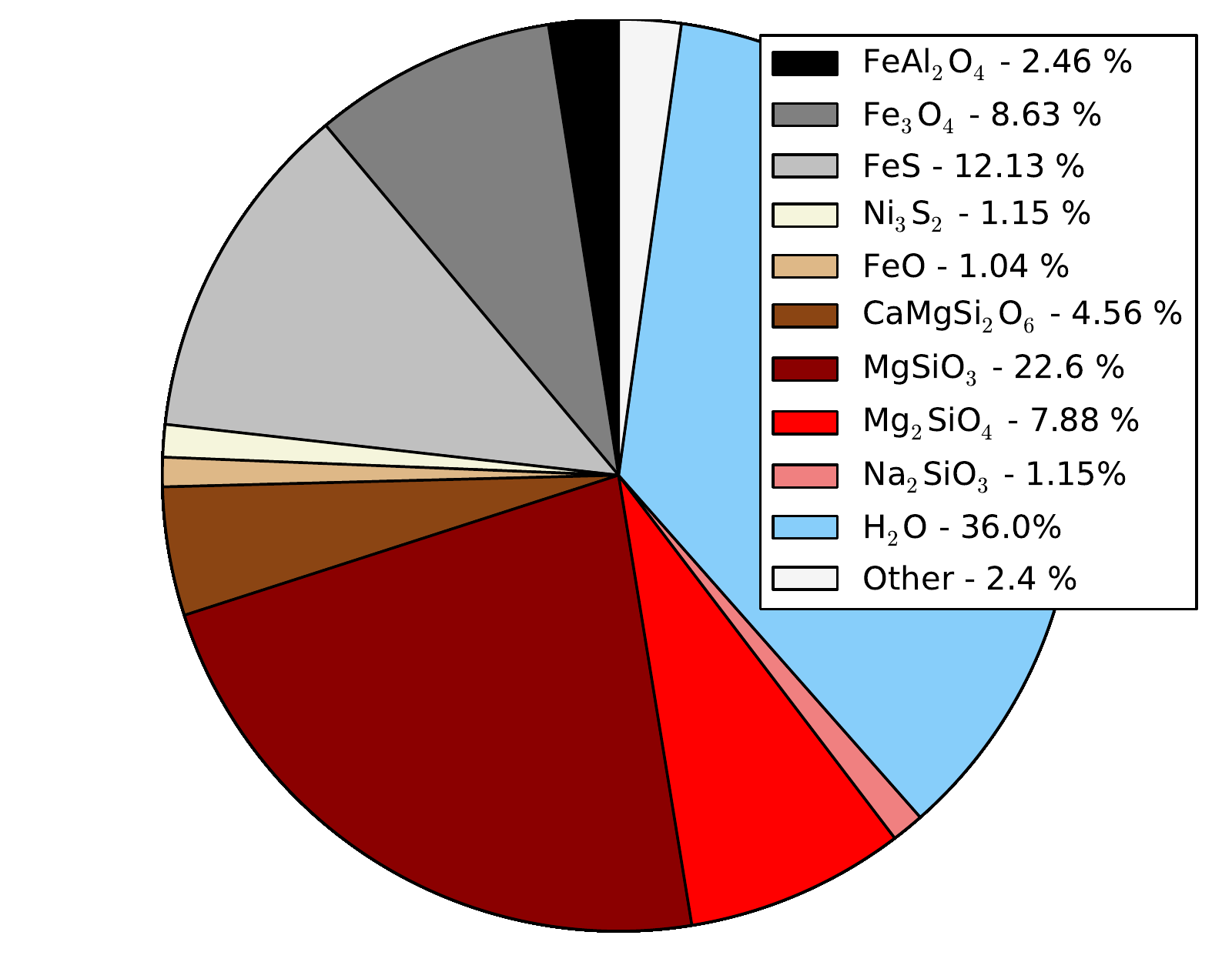}
\label{fig:14a}
}
\subfloat[Ternary diagrams showing results of planet interior structure calculations. The radius (shown in units of km) of a super Earth is dependent on the planet's total mass, as well as the mass distribution among water, core (irons and nickels), and mantle (Mg/Al/Ca silicates) components. Figure from \citet{Valencia2007}, ApJ, 665, 1413. Reproduced with permission \textcopyright AAS.]{
\includegraphics[width=0.5\textwidth]{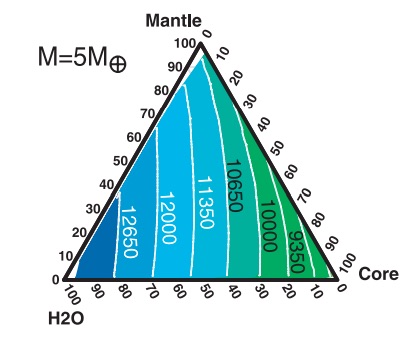}
\label{fig:14b}
}
\caption{The link between a super-Earth's size and its internal composition is complicated, but generally dominated by the abundance of ice that it accretes and retains during its formation.}
\label{fig:14}
\end{figure*}

Planets will acquire their rocky materials while migrating through the disk in planet traps, accreting materials characteristic
of those traps (eg. ice lines). The movement of 
the trap through the evolving disk determines the cumulative inventory of all of the different kinds
of solid materials that are accreted along the way \citep{APC16a}.   Figure 8.a shows the detailed breakdown of materials accreted from the
disk by a forming super Earth planet moving with the heat transition trap.   The final mass in this model reaches 5.4 Earth masses.
In this calculation, the ice fraction is 36 \%,  the mantle 36 \%, and core materials 27 \%.  Since 
cores trapped at
the heat transition will spend most of their life beyond the water ice line, these will generally have the greatest ice mass fractions.  
Planets trapped at the water ice line will have a smaller proportion of ice, while dead zone planets will have the least since the dead zone radius
occurs typically inside the ice line.   As an example of the sensitivity of the compositions of Super Earth compositions to disk parameters, planets formed in a disk with a 2 Myr life time
in he heat transition planet had an ice content reached 48 \%, whereas for the dead zone planets only reached 6 \%.   

As was the case for solids, gas abundances of planet atmospheres will reflect the temperature of the regions in the disk where they accreted material. Abundances of CO, N$_2$ and SiO result from gas accretion in hot regions
of the disc, while accretion from colder regions of the disc results
in higher abundances of H$_2$, O, CH$_4$, and NH$_3$. 
These results clearly illustrate
how a wide range of bulk densities of planets can arise when one considers variations across disk populations (eg. 
different lifetimes), as well as the type of traps planets accrete their materials from.  
This gives some explanation for the wide scatter seen in the M-R diagram.  

The materials delivered to a planet during the accretion phase are of course modified as a consequence of the P-T relations
in the planet interior and atmosphere.  
Terrestrial planets are modelled as having a crust, mantle, liquid core, and a solid core.  The most massive of these are the mantle and 
the core, comprised of silicates and iron alloys.  These differentiated from one another because iron is denser than silicates.  Four elements
(oxygen, iron, magnesium and silicon) account for 95 \% of the total mass of the Earth  \citep{Javoy1995} .     

The M-R relation 
for rocky material at constant density is $ R \propto M^{1/3} $.  
The dependence of the M-R relation on the types of materials in the planet can be  
simplified by considering just 
3 basic types of materials;  ice,  mantle materials, and iron (the latter two being categories 
containing many minerals).   Following the early work on zero temperature models  for single
compositions by \cite{ZapolskySalpeter1969}, \cite{Valencia2007} built models that include all
three of these basic materials.  Here, the P-T relation diagram is computed using EOS used for modelling 
the Earth.  The results of such calculations are illustrated in terms of ternary diagrams often used in Earth sciences 
(Figure \ref{fig:14b} ).   Data for this three component system are plotted along the sides of a triangle, representing 
each of these basic materials.  Each vertex means 100 \% of a particular component, and data
plotted on the opposite side mean 0 \%. Lines parallel to a particular
side will show various degrees of a component whose maximum
value is shown at the corresponding opposite vertex.   The M-R relationship is found to
take a power law form  $R \propto M^{\beta}$ where $\beta = 0.262 ( 1.0 - 0.3 IMF ) $ and  IMF is the
percentage amount of water.  For super Earths, pressures will be much higher and so different EOS are required.
This M-R power law exponent is actually less than $ 1/3$ (0.274) because as 
the density increases with increasing pressure while the temperature remains roughly constant \citep{Grasset2009}.   

\subsection{Atmospheres} 

A key observable in studying the chemical composition of exoplanetary atmospheres is the carbon-to-oxygen ratio (C/O). This is because (as we have said before) it can be naturally linked to chemical processes in the protoplanetary disk, and also because the elemental ratio is less sensitive to chemical processes within the atmosphere.

\begin{figure*}
\centering
\includegraphics[width=\textwidth]{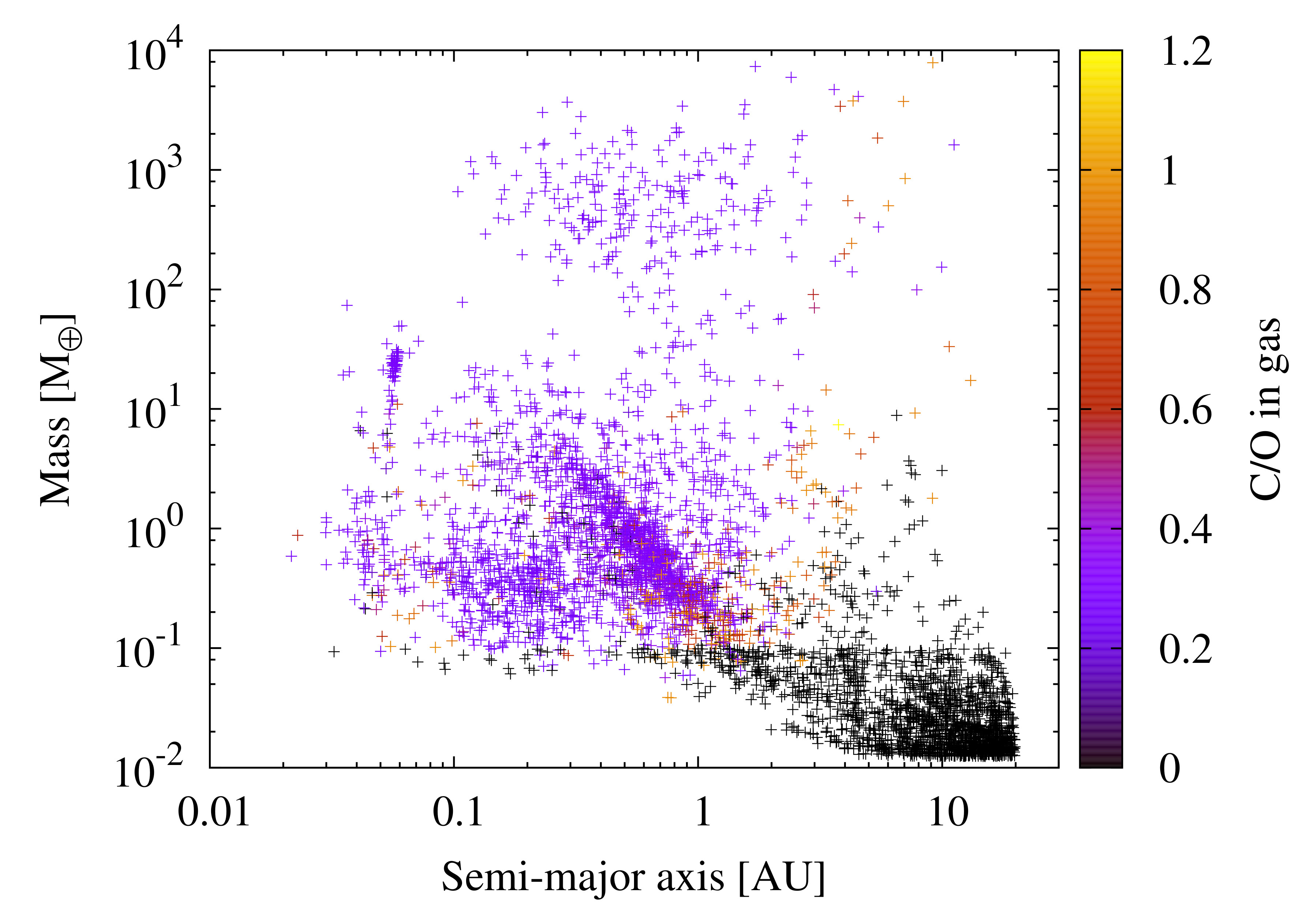}
\caption{Distribution of C/O in the atmosphere of planets across a population synthesis model. In this model the molecular abundances are held constant, and hence the C/O of the planetary atmosphere is generally dominated by the accretion of gas. Figure from \citet{Thiabaud2015}, A\&A 574, A138. Reproduced with permission \textcopyright ESO.}
\label{fig:08}
\end{figure*}

In Figure \ref{fig:08} we show a population synthesis model that included enough simple chemistry to estimate the resulting C/O (from \cite{Thiabaud2015}). In their work, \cite{Thiabaud2015} showed that the resulting C/O in a planetary atmosphere is dependent on the chemical evolution of the disk, the migration and accretion history of the planet, and the modelled physical structure of the protoplanetary disk. In the figure, the radial distribution of the disk's molecular abundances is held constant throughout planet formation, similar to the assumptions made to construct \ref{fig:06} by \cite{Oberg11}. However, \cite{Thiabaud2015} also include a model where the gas diffuses inward as the host star accretes material through the disk. In this model, the planetary C/O is dominated by the accretion of ices, this is discussed below.

\cite{Thiabaud2015} initialize their disk model with C/O that is approximately solar (0.54).  Generally it is assumed that the disk's initial C/O (in both gases and solids) matches the stellar C/O that can be observed today. They find that most planets have sub-stellar C/O ($<0.3$), while some of their planets resulted in super-stellar C/O. This result depended on their choice of chemical model, and in the case that ice accretion dominated the C/O, nearly all planets had sub-stellar C/O. The majority of their high mass, close-in planets (ie. Hot Jupiters) generally have sub-stellar C/O which disagrees with the observations of \cite{Brewer16}.

A similar conclusion was reached by \cite{Mordasini16}, and as in the case of the majority of models run by \cite{Thiabaud2015}, C/O is dominated by the accretion of ices. In \cite{Crid17} C/O is dominated by the accretion of the gas, and because of their migration model, most of the planets accrete their gas near the water ice line, resulting in planetary C/O which matched the initial gas C/O of the disk. A key difference between models which are dominated by gas and ice accretion is in their treatment of the chemical abundance of the gas. 

In the evolving models of \cite{Thiabaud2015} and in \cite{Mordasini16} the gas is assumed to be `pristine' - only consisting of H/He - while the higher metalicity gas is accreted onto the host star.   However,  \cite{Crid17} compute the chemical evolution of the gas while ignoring the radial accretion of the gas through the disk. Observations of protoplanetary disk \citep{HenningSemenov2013}, along with simulations which include chemistry, gas evolution, and dust radial drift \citep{Bosman2017}, tend to disfavour pristine gas in the planet forming region of the disk.

\begin{figure*}
\centering
\subfloat[Planet formation tracks for three planets formed in the water ice line (blue), dead zone (black) and heat transition (orange). The points on each track denote the position of the growing planet a 1,2,3, and 4 Myr. The dotted lines denote the location of the water ice line at each of the labels times.]{
\includegraphics[width=0.5\textwidth]{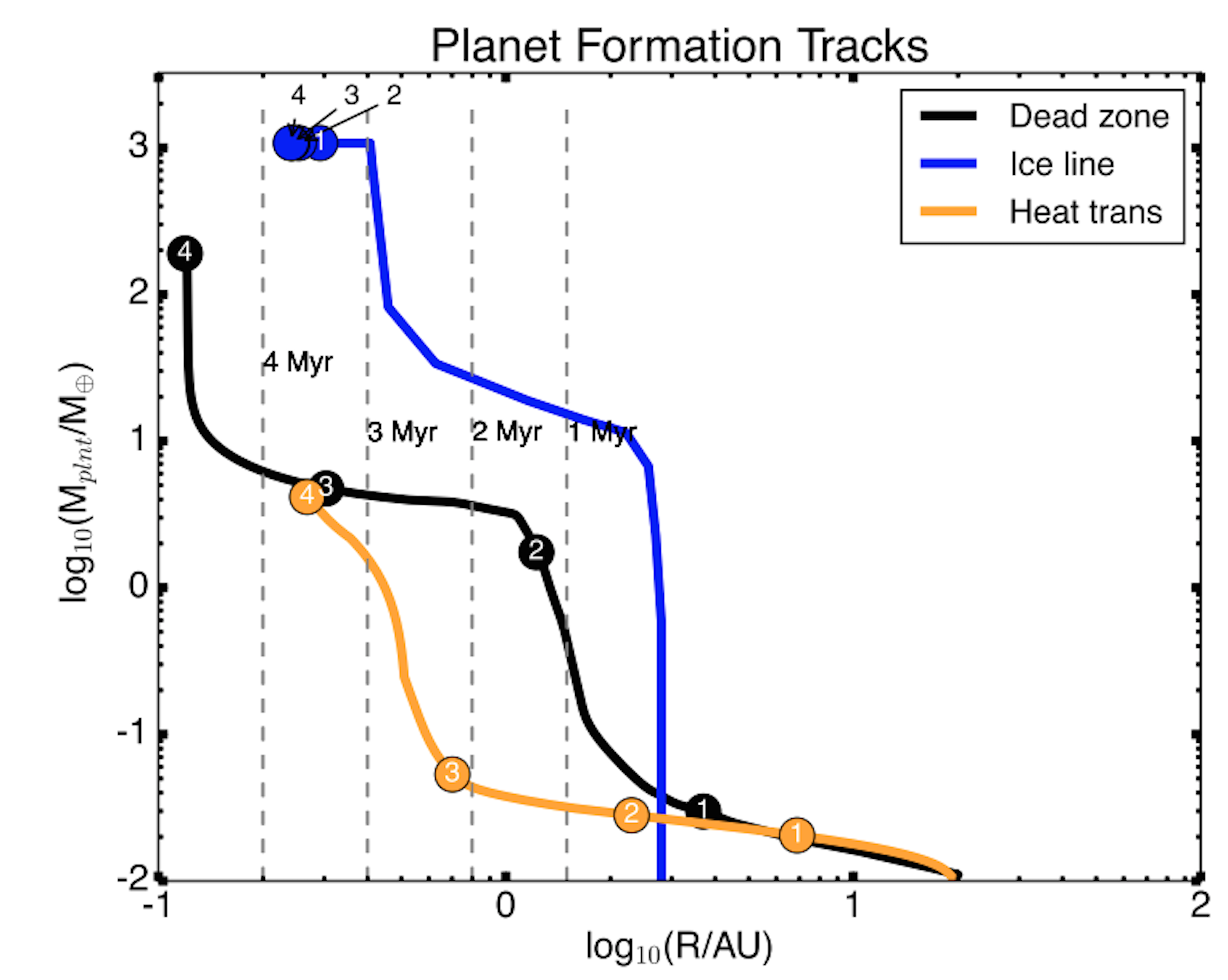}
\label{fig:08xa}
}
\subfloat[Gas abundance of the `minor gases' - most abundant gases other than Hydrogen and Helium. While the CO and H$_2$O abundances are nearly identical across the three planets, the dominant nitrogen carrier (N$_2$/NH$_3$) depend on {\it when} the gas accretes. Early: NH$_3$, Late: N$_2$.]{
\includegraphics[width=0.5\textwidth]{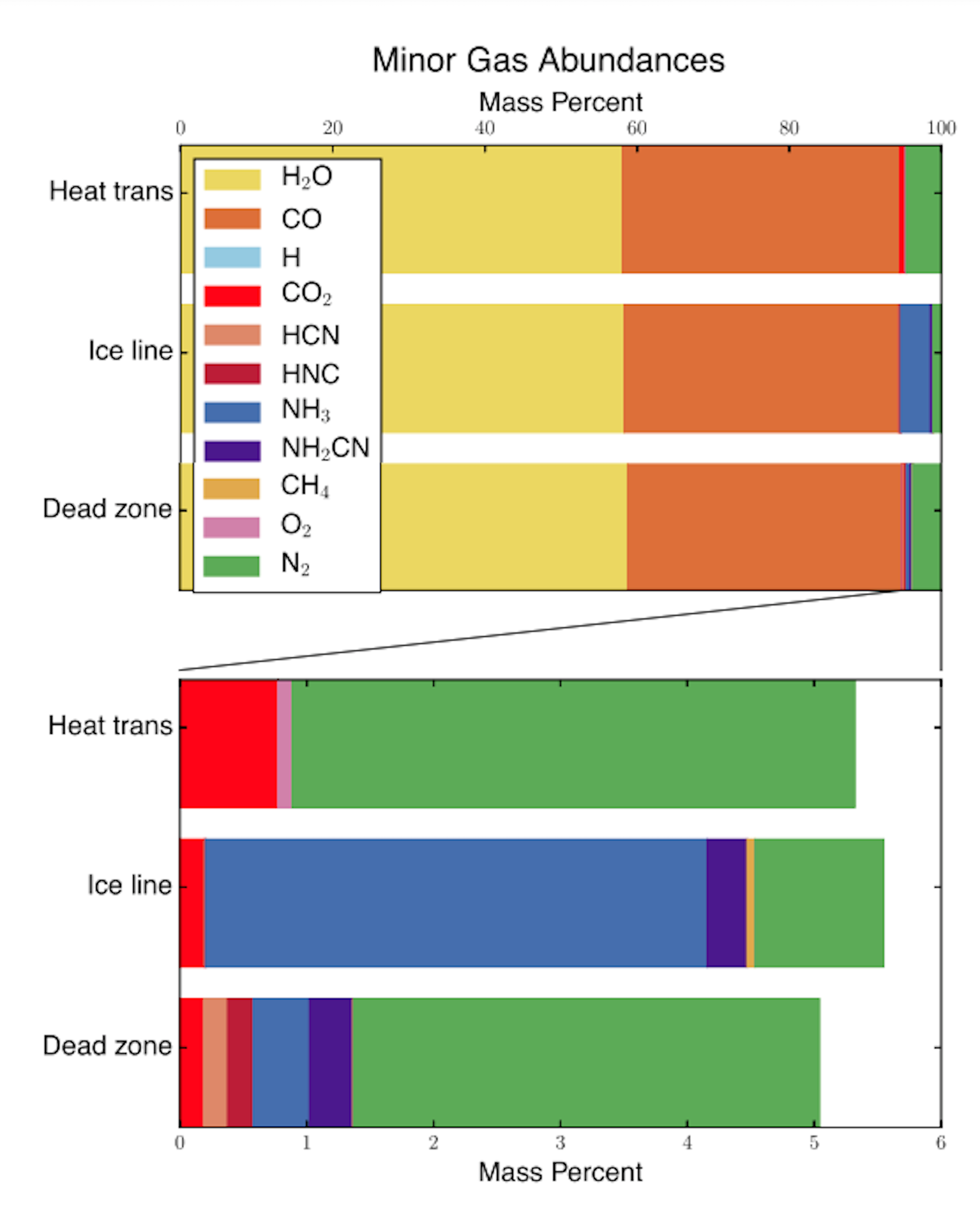}
\label{fig:08xb}
}
\caption{Molecular abundance results for planet formation in an astrochemically evolving disk model. Here we see that {\it when} a planet accretes its gas can be as important as {\it where} it accretes because the molecular gas can undergo chemical evolution on a similar timescale as the planet formation process. Figures reproduced from Cridland et al. (2017), MNRAS, 469, 3910.}
\label{fig:08x}
\end{figure*}

In Figure \ref{fig:08xa} we show a set of planet tracks used in \cite{Crid17}. Each track denotes the evolution of a planet in each individual planet trap, and we label the time (in Myr) when the planet appears at a given position. Along side these tracks we note the location of the water ice line with vertical dotted lines at 1-4 Myr. 

In Figure \ref{fig:08xb} we show the molecular abundance of the atmospheres accreted by the planets  from the left panel. While the C/O and the C/N (carbon-to-nitrogen ratio) are the same in each of these planetary atmospheres, their initial molecular abundance can differ. In particular \cite{Crid17} report a difference in the abundance of the primary nitrogen carriers (N$_2$ and NH$_3$). The planet which accreted its gas early (the ice line trapped planet) primarily accreted NH$_3$ while the two planets that accreted later in the disk evolution accreted primarily N$_2$. This highlights an important additional restriction on chemical composition of planetary atmosphere. Because of chemical evolution in the protoplanetary disk, {\it when} and {\it where} a planets accretes its gas will impact the molecular abundance that is inherited by the planet's atmosphere.

Further complicating this problem are the effects that enrichment from solids can have on a planet's atmospheric composition. There are two important effects to consider that can lead to atmospheric enrichment from solids: (i) Solids accreted from the disk onto a planet with an atmosphere can be disrupted or ablated \citep{Pinhas2016} instead of accreting directly onto the core. This effect has also been recently considered to investigate the maximum masses of cores that can be built up by accretion of pebbles \citep{Alibert2017, Brouwers2017}. (ii) As was discussed in \citet{Madhusudhan2017}, a planetary core can be eroded by the massive atmosphere of a gas giant. Convective motions throughout a Jovian atmosphere can mix the eroded, metal-rich material throughout the atmosphere, changing its composition and C/O ratio. These effects suggest that that accretion rates and compositions of both gases and solids need to be considered to fully understand the compositions of planet atmospheres throughout the formation phase.  

An important result overall is that intermediate size, low-velocity planetesimals of
90 - 250 metres  can penetrate through a massive envelope and reach the core. These objects
are too small to attain the velocities required for frictional
ablation and yet too large to be heated to the melting temperature. The key question then becomes - at what point
in the growth of the atmosphere and migration of the planet were conditions most favourable for planetesimal accretion to the core, and what fraction of the 
overall solid accretion was this? The answer to this question clearly involves a much better understanding of coupled migration and accretion.     

Once in the atmosphere the gas can evolve both physically and chemically between when it was accreted and when it is observed. These processes are complex, and their study involve three-dimensional hydrodynamic simulations (eg. \cite{Cooper2006}), time-dependent photo-chemical networks (eg. \cite{Agundez2014}), and synthetic spectra (eg. \cite{Molliere2017}). As it stands, no complete model of exoplanetary atmospheres has been developed, and the observations of chemical complexity in atmosphere remain in their early stages of development. With the next generation of space and ground based telescopes we will soon see exoplanetary atmospheres in a new light.

\section{Conclusions}

Having reviewed the basic processes involved in an end-to-end picture of planet formation, we return to our original question.  Are there clearly discernible links between the chemical and physical structure of planets, and their formation history?    We have seen that on the observational side, JWST is likely to revolutionize our understanding of the composition of 
planetary atmospheres.   On the theoretical side, news ideas related to angular momentum transport by MHD disk winds as well as the role of pebble accretion are still very much in development.  And finally, the Juno results which have shed so much new light to the problem of Jupiter's structure and origin are still being derived.  We are likely to be on the cusp of a radically new picture of planet formation and composition as these results solidify over the coming 5 years.   Be that as it may, one can still discern some important emerging patterns. 

\begin{itemize}
\item  {\bf The core accretion picture has ever more empirical support. }  
The results of the Juno probe and the ability of such models to understand the mass-metallicity relation and many aspects of observed
planetary populations  appear to give this picture a growing amount of empirical and theoretical support.   

\item {\bf Migration and traps have a direct effect on the kinds of materials that planets accrete. } 
The fact that planets accrete most of their solids while in particular kinds of traps gives them special kinds of "diets" that are preserved in the element ratios
of their cores and atmospheres.  More work is needed to perfect our ideas of traps and migration, but because these are connected to both disk chemistry 
and angular momentum transport, element patterns arising from accretion probably persist.  

\item {\bf Dust evolution is critical.}
The movement of dust in protoplanetary disks affects all aspects of disk ionization, disk chemistry, and ultimately on the nature of the solids that are accreted
by growing, migrating planets.  We do not yet have a final picture of dust evolution in disks, but observations from ALMA and JWST may help considerably in
building more comprehensive physical theories of dust evolution in disks. 

\item  {\bf  Accretion vs dissolution in massive planet interiors }.    The Juno results underline some of the important and lingering uncertainties in theoretical modelling of the formation of massive planets.   Jupiter's core could be slowly dissolving since its formation, or, planetesimals may not have reached the core and dissolved instead in the envelope.  
The pattern of chemical enrichment will be different in these two scenarios so greater effort needs to be applied to understanding the detailed fates of planetesimals as they accrete into the  atmospheres of giant planets during formation. 

\end{itemize}

While no clear cut answer to the question is possible at this point, there are reasons to expect that these links do indeed exist.   Elucidating them will require the
best new instruments, observatories, and calculations that we can muster over the next decade.    

\section{Acknowledgements}  We thank Phil Armitage for his very useful referee report.  We also thank Yasuhiro Hasegawa, Ted Bergin, Til Birnstiel, Christoph Mordasini, Thomas Henning, Dimitry Semenov, Nikku Madhusudhan, Richard Nelson, and Colin McNally for enlightening discussions during the course of this project.   This research was supported by a Discovery grant to REP from the Natural Sciences and Engineering Research Council of Canada (NSERC) , as well as by NSERC postgraduate scholarships to AC and MA.

\bibliographystyle{spbasicHBexo}  
\bibliography{mybib} 

\end{document}